\documentclass[]{mn2e}
\usepackage{epsfig}
\def\LB{\hbox{$\lambda$ Bo{\"o}tis }}
\title{The status of Galactic field $\lambda$ Bo{\"o}tis stars in the post-Hipparcos era}
\author[E.~Paunzen, I.Kh.~Iliev, I.~Kamp, I.S.~Barzova]
{E.~Paunzen$^{1,2}$, I.Kh.~Iliev$^{3}$, I.~Kamp$^{4}$, I.S.~Barzova$^{3}$ \\
1-Institut f\"{u}r Astronomie der Universit\"{a}t Wien, T\"{u}rkenschanzstr.
17, 1180 Wien, Austria \\
2-Zentraler Informatikdienst der Universit\"{a}t Wien,
Universit\"{a}tsstr. 7, 1010 Wien, Austria \\
3-Institute of Astronomy, National Astronomical Observatory,
P.O. Box 136, 4700 Smolyan, Bulgaria \\
4-Leiden Observatory,
Niels Bohrweg 2, PO Box 9513, 2330 RA Leiden, The Netherlands
}
\pubyear{}
\begin{document}
\maketitle
\begin{abstract}
The \LB stars are Population\,I, late B to early F-type stars, 
with moderate to extreme (up to a factor 100) surface underabundances of most
Fe-peak elements and solar abundances of lighter elements (C, N, O, and S).
To put constraints on the various existing theories that try to explain 
these peculiar stars, we investigate the observational properties of 
\LB stars compared to a reference sample of normal stars. 
Using various photometric systems and Hipparcos data, we analyze the 
validity of standard photometric calibrations, elemental abundances, 
and Galactic space motions.

There crystallizes a clear picture of a homogeneous group of Population\,I
objects found at all stages of their main-sequence evolution, with a peak at about
1~Gyr. No correlation of astrophysical parameters such as the projected
rotational velocities or elemental abundances with age is found,
suggesting that the a-priori unknown mechanism, which creates \LB 
stars, works continuously for late B to early F-type stars in all stages of 
main-sequence evolution. Surprisingly, the sodium abundances seem to indicate
an interaction between the stars and their local environment.

\end{abstract}
\section{Introduction}

Knowledge of the evolutionary status of the members of the
\LB group is essential to put tight constraints on the astrophysical
processes behind this phenomenon. We have chosen the following working definition
as group characteristics: 
late B- to early F-type, Population\,I stars with apparently solar abundances 
of the light elements (C, N, O and S) and moderate to strong underabundances of 
Fe-peak elements (see Faraggiana \& Gerbaldi 1998 for a critical
summary of the various definitions). Only a maximum of about 2\% of all objects in the
relevant spectral domain are believed to be \LB\hskip-2pt-type stars (Paunzen
2001). That already suggests either
that the mechanism responsible for the phenomenon
works on a very short time-scale 
(10$^{6}$\,yrs) or else the general conditions for the development are very strict.

We know already of a few \LB stars in the Orion OB1 association
and one candidate in NGC~2264 (Paunzen 2001), for both of which log\,$t$\,$\approx$\,7.0.
The evolutionary status for two \LB\hskip-2pt-type spectroscopic-binary systems 
(HD~84948 and HD~171948) 
has been determined as very close to the Zero Age Main Sequence (ZAMS hereafter)
for HD~171948 and to the Terminal Age Main Sequence (TAMS hereafter) for HD~84948 (Iliev et
al. 2002). The results for the other Galactic field stars are not clear.

In 1995, Iliev \& Barzova summarized the evolutionary status of 20
\LB stars (and Vega) using Str\"omgren $uvby\beta$ photometric
data. They concluded that most of the stars
studied are in the middle of their main-sequence evolution, with only a
few objects near the ZAMS.

Paunzen (1997) investigated the parallaxes measured by the Hipparcos satellite
for a sample of \LB\hskip-2pt-type stars in order to derive luminosities, masses and
ages for 18 objects in common with Iliev \& Barzova (1995). He found no systematic
influence of the distance, effective temperature, metallicity and rotational
velocity on the difference between photometrically calibrated absolute
magnitudes and those derived from Hipparcos parallaxes. 
Six objects were found to be very close
to the ZAMS and a hypothesis was proposed that all other stars are in their
Pre-Main-Sequence (PMS hereafter) phase.

Later on, Bohlender, Gonzalez \& Matthews (1999) and Faraggiana \& Bonifacio (1999)
challenged that hypothesis with plausible arguments such as the
unusually vigorous star-forming activity that it implied
in the solar neighbourhood and a statistical analysis of normal-type stars. 

Already Gray \& Corbally (1998) stated that the \LB phenomenon can
be found from very early stages to well into the main-sequence
life of A-type stars. That conclusion was based on the incidence
of \LB stars among very young A-type stars, which is not very different
from the incidence among Galactic field stars.

In this paper we present a much more extensive investigation,
including the data of the Hipparcos satellite. With the help of
photometric data of the Johnson {\it UBV}, Str\"omgren $uvby\beta$
and Geneva 7-colour systems, different calibrations of the absolute
magnitude, effective temperature and surface gravity are applied
and compared.

From evolutionary models (Claret 1995), masses and ages
are estimated. They are compared with those derived by
Iliev \& Barzova (1995) and Paunzen (1997). 

As a further step, the proper motions of \LB stars are used
to calculate space velocities. That very important information
should help further to sharpen the group properties and to
sort out probably misclassified stars. 

Another point investigated is the question as to whether there exists a typical
abundance pattern for the \LB group. Heiter (2002) and
Heiter et al.~(2002) tried to shed more light on the abundance pattern
in the context of the proposed theories. We have searched for correlations
of the individual abundances with, especially, mass and age.

However, we will not comment on our results in the context of the 
developed theories and models, because they still depend on too many free
parameters. The aim of this paper is not to promote
one of the suggested theories but rather to find strict observational constraints which
should be incorporated into future theoretical investigations. 
\def\b{\hskip0.5em}  \def\bb{\hskip1em}  \def\d{\hskip-0.7em}
\begin{table*}
\caption[]{Photometric data, stellar parameters and calibrated values for
the program stars. In
parenthesis are the errors in the final digits of the corresponding quantity.} \label{lb_basic}
\begin{center}
\begin{tabular}{lllccccrllll}
\hline
\bb HD	&	\b HR & \b HIP	&	$V$	&	$B-V$	&	$b-y$	&	$A_{\rm V}$	&	\multicolumn{1}{c}{$v$\,sin\,$i$} & 
\multicolumn{1}{c}{$T_{\rm eff}$}		&	\multicolumn{1}{c}{\d log\,$g$(phot)}	&	\multicolumn{1}{c}{$M_{\rm V}$}		&	
\multicolumn{1}{c}{\d log\,L$_{\ast}$/L$_{\odot}$}	\\
& & & \multicolumn{1}{c}{[mag]} & \multicolumn{1}{c}{[mag]} & \multicolumn{1}{c}{[mag]} & \multicolumn{1}{c}{[mag]} 
& \multicolumn{1}{c}{[kms$^{-1}$]} & \multicolumn{1}{c}{[K]} & \multicolumn{1}{c}{\d [dex]} & 
\multicolumn{1}{c}{[mag]} \\
\noalign{\vskip6pt}
\bb\b319	&\bb	12&\bb\b636	&	5.934	&	0.141	&	0.079	&	0.004	&{60\bb}	&	8020(135)	&	3.74(8)	&	1.27(19)	&	1.45(8)	\\
\bb6870	&		&	\bb5321	&	7.494	&	0.246	&	0.164	&	0.000	&{165\bb}	&	7330(102)	&	3.84(11)	&	2.29(42)	&	1.02(17)	\\
\bb7908	&		&	\bb6108	&	7.288	&	0.272	&	0.192	&	0.000	&		&	7145(87)	&	4.10(12)	&	2.60(18)	&	0.90(7)	\\
\b11413	&	 \b 541	&	\bb8593	&	5.940	&	0.147	&	0.105	&	0.004	&{125\bb}	&	7925(124)	&	3.91(21)	&	1.49(10)	&	1.35(4)	\\
\b13755	&		&	\b10304	&	7.844	&	0.318	&	0.181	&	0.000	&		&	7080(161)	&	3.26(10)	&	0.93(10)	&	1.57(4)	\\
\b15165	&		&	\b11390	&	6.705	&	0.333	&	0.191	&	0.010	&{90\bb}	&	7010(167)	&	3.23(10)	&	1.12(16)	&	1.50(6)	\\
\b23392	&		&	\b17462	&	8.260	&	0.020	&	0.014	&	0.094	&		&	9805(281)	&	4.35(9)	&	1.43(30)	&	1.45(12)	\\
\b24472	&		&	\b18153	&	7.092	&	0.304	&	0.214	&	0.003	&		&	6945(131)	&	3.81(16)	&	2.14(11)	&	1.09(5)	\\
\b30422	&	1525	&	\b22192	&	6.186	&	0.190	&	0.101	&	0.014	&{135\bb}	&	7865(108)	&	4.00(20)	&	2.35(1)	&	1.01(1)	\\
\b31295	&	1570	&	\b22845	&	4.648	&	0.085	&	0.044	&	0.063	&{115\bb}	&	8920(177)	&	4.20(1)	&	1.66(22)	&	1.32(9)	\\
\b35242	&	1777	&	\b25205	&	6.348	&	0.122	&	0.068	&	0.042	&{90\bb}	&	8250(103)	&	3.90(14)	&	1.75(22)	&	1.26(9)	\\
\b54272	&		&		&	8.800	&	0.261	&	0.214	&	0.000	&		&	7010(217)	&	3.83(10)	&	2.33(30)	&	1.01(12)	\\
\b74873	&	3481	&	\b43121	&	5.890	&	0.115	&	0.064	&	0.078	&{130\bb}	&	8700(245)	&	4.21(11)	&	1.82(1)	&	1.24(1)	\\
\b75654	&	3517	&	\b43354	&	6.384	&	0.242	&	0.161	&	0.012	&{45\bb}	&	7350(104)	&	3.77(11)	&	1.83(12)	&	1.20(5)	\\
\b81290	&		&	\b46011	&	8.866	&	0.332	&	0.254	&	0.124	&{55\bb}	&	6895(214)	&	3.82(28)	&	1.85(30)	&	1.20(12)	\\
\b83041	&		&	\b47018	&	8.927	&	0.294	&	0.223	&	0.161	&{95\bb}	&	7120(208)	&	3.76(20)	&	1.70(30)	&	1.26(12)	\\
\b83277	&		&	\b47155	&	8.304	&	0.311	&	0.226	&	0.131	&		&	7000(189)	&	3.67(18)	&	1.49(29)	&	1.35(12)	\\
\b84123	&		&	\b47752	&	6.840	&	0.297	&	0.235	&	0.040	&{20\bb}	&	7025(175)	&	3.73(17)	&	1.58(15)	&	1.31(6)	\\
\b87271	&		&	\b49328	&	7.120	&	0.172	&	0.151	&	0.008	&		&	7515(232)	&	3.43(10)	&	1.02(8)	&	1.53(3)	\\
\b90821	&		&		&	9.470	&	0.105	&	0.068	&	0.013	&{150\bb}	&	8190(79)	&	3.73(10)	&	0.74(30)	&	1.66(12)	\\
\b91130	&	4124	&	\b51556	&	5.902	&	0.109	&	0.073	&	0.000	&{135\bb}	&	8135(98)	&	3.78(10)	&	1.36(26)	&	1.42(11)	\\
101108	&		&	\b56768	&	8.880	&	0.179	&	0.114	&	0.006	&{90\bb}	&	7810(80)	&	3.90(18)	&	1.33(30)	&	1.42(12)	\\
102541	&		&	\b57567	&	7.939	&	0.230	&	0.163	&	0.097	&		&	7665(168)	&	4.22(16)	&	2.34(21)	&	1.01(9)	\\
105058	&		&	\b58992	&	8.900	&	0.183	&	0.129	&	0.009	&{140\bb}	&	7740(171)	&	3.77(30)	&	0.86(30)	&	1.60(12)	\\
105759	&		&	\b59346	&	6.550	&	0.218	&	0.142	&	0.000	&{120\bb}	&	7485(102)	&	3.65(10)	&	1.35(21)	&	1.40(8)	\\
106223	&		&	\b59594	&	7.431	&	0.288	&	0.228	&	0.015	&{90\bb}	&	6855(247)	&	3.49(18)	&	1.83(45)	&	1.22(18)	\\
107233	&		&	\b60134	&	7.353	&	0.255	&       0.192	&	0.048	&{95\bb}  &	7265(143)	&	4.03(10)	&	2.64(13)	&	0.88(5)	\\
109738	&		&		&	8.277	&	0.198	&	0.161	&	0.073	&		&	7610(145)	&	3.90(13)	&	1.85(30)	&	1.20(12)	\\
110377	&	4824	&	\b61937	&	6.228	&	0.195	&	0.120	&	0.000	&{170\bb}	&	7720(89)	&	3.97(14)	&	1.96(11)	&	1.16(5)	\\
110411	&	4828	&	\b61960	&	4.881	&	0.077	&	0.040	&	0.045	&{165\bb}	&	8930(206)	&	4.14(14)	&	1.90(28)	&	1.22(11)	\\
111005	&		&	\b62318	&	7.959	&	0.376	&	0.224	&	0.009	&		&	6860(66)	&	3.72(10)	&	1.76(53)	&	1.24(21)	\\
111604	&	4875	&	\b62641	&	5.886	&	0.160	&	0.112	&	0.000	&{180\bb}	&	7760(149)	&	3.61(25)	&	0.48(7)	&	1.75(3)	\\
120500	&		&	\b67481	&	6.600	&	0.131	&	0.068	&	0.017	&{125\bb}	&	8220(70)	&	3.86(10)	&	0.85(34)	&	1.62(13)	\\
120896	&		&	\b67705	&	8.495	&	0.296	&	0.166	&	0.000	&		&	7260(89)	&	3.76(10)	&	1.90(30)	&	1.18(12)	\\
125162	&	5351	&	\b69732	&	4.186	&	0.084	&	0.051	&	0.039	&{115\bb}	&	8720(156)	&	4.07(9)	&	1.71(23)	&	1.28(9)	\\
125889	&		&		&	9.849	&	0.241	&	0.206	&	0.108	&		&	7275(175)	&	3.88(9)	&	2.32(30)	&	1.01(12)	\\
130767	&		&	\b72505	&	6.905	&	0.042	&	0.002	&	0.000	&		&	9195(220)	&	4.10(8)	&	1.27(1)	&	1.48(1)	\\
142703	&	5930	&	\b78078	&	6.120	&	0.240	&	0.182	&	0.021	&{100\bb}	&	7265(150)	&	3.93(12)	&	2.41(12)	&	0.97(5)	\\
142944	&		&		&	7.179	&	0.293	&	0.201	&	0.013	&{180\bb}	&	7000(125)	&	3.19(4)	&	0.80(30)	&	1.62(12)	\\
149130	&		&	\b81329	&	8.498	&	0.342	&	0.233	&	0.109	&		&	6945(103)	&	3.49(5)	&	1.51(30)	&	1.34(12)	\\
153747	&		&	\b83410	&	7.420	&	0.140	&	0.098	&	0.128	&		&	8205(90)	&	3.70(24)	&	1.24(30)	&	1.46(12)	\\
154153	&	6338	&	\b83650	&	6.185	&	0.284	&	0.199	&	0.020	&		&	7055(120)	&	3.56(6)	&	1.86(29)	&	1.19(11)	\\
156954	&		&	\b84895	&	7.679	&	0.294	&	0.200	&	0.050	&{50\bb}	&	7130(93)	&	4.04(13)	&	2.81(33)	&	0.82(13)	\\
168740	&	6871	&	\b90304	&	6.138	&	0.201	&	0.136	&	0.035	&{145\bb}	&	7630(81)	&	3.88(14)	&	1.82(2)	&	1.21(1)	\\
168947	&		&		&	8.123	&	0.196	&	0.172	&	0.116	&		&	7555(185)	&	3.67(10)	&	1.28(30)	&	1.43(12)	\\
170680	&	6944	&	\b90806	&	5.132	&	0.008	&	0.008	&	0.091	&{205\bb}	&	9840(248)	&	4.15(6)	&	0.83(23)	&	1.70(9)	\\
175445	&		&	\b92884	&	7.792	&	0.110	&	0.055	&	0.108	&		&	8520(198)	&	3.96(10)	&	1.08(27)	&	1.53(11)	\\
183324	&	7400	&	\b95793	&	5.795	&	0.084	&	0.051	&	0.083	&{90\bb}	&	8950(204)	&	4.13(4)	&	1.64(42)	&	1.32(17)	\\
184779	&		&		&	8.940	&	0.224	&	0.187	&	0.000	&		&	7210(173)	&	3.63(21)	&	1.26(30)	&	1.43(12)	\\
192640	&	7736	&	\b99770	&	4.934	&	0.154	&	0.099	&	0.016	&{80\bb}	&	7940(96)	&	3.95(18)	&	1.84(1)	&	1.22(1)	\\
193256	&	7764C	&	100286	&	7.721	&	0.213	&	0.116	&	0.063	&{250\bb}	&	7740(94)	&	3.69(17)	&	1.08(30)	&	1.51(12)	\\
193281	&	7764A	&	100288	&	6.557	&	0.190	&	0.098	&	0.111	&{95\bb}	&	8035(115)	&	3.54(4)	&	0.41(30)	&	1.79(12)	\\
198160	&	7959	&	102962	&	5.663	&	0.189	&	0.108	&	0.022	&{200\bb}	&	7870(129)	&	3.99(9)	&	1.47(41)	&	1.36(16)	\\
204041	&	8203	&	105819	&	6.456	&	0.161	&	0.092	&	0.026	&{65\bb}	&	7980(97)	&	3.97(8)	&	1.75(18)	&	1.25(7)	\\
210111	&	8437	&	109306	&	6.377	&	0.203	&	0.136	&	0.000	&{55\bb}	&	7550(123)	&	3.84(15)	&	1.76(15)	&	1.23(6)	\\
216847	&		&	113351	&	7.060	&	0.242	&	0.155	&	0.000	&		&	7355(78)	&	3.47(14)	&	0.93(24)	&	1.56(10)	\\
221756	&	8947	&	116354	&	5.576	&	0.095	&	0.056	&	0.043	&{105\bb}	&	8510(188)	&	3.90(3)	&	1.16(16)	&	1.50(6)	\\
\hline
\end{tabular}
\end{center}
\end{table*}

\section{Program stars and their astrophysical parameters}

The program stars were taken from the lists of Gray \& Corbally (1998)
and Paunzen (2001), with the omission of apparent spectroscopic binary
systems (e.g.~HD~38545, HD~64491, HD~111786, HD~141851 and HD~148638) 
as well as of HD~191850
for which no $\beta$ measurement is available so no reliable
astrophysical parameters could be derived from the Str\"omgren photometric system. A further critical 
assessment
of the literature was performed in order to reject probable non-members
and ill-defined objects. In total, 57 well established \LB stars were chosen. 

The photometric data were taken from the General Catalogue of Photometric Data (GCPD;
http://obswww.unige.ch/gcpd/) as well as from the Hipparcos and Tycho
database. If available, averaged and weighted mean values were used.

The following calibrations for the individual photometric systems were
used to derive the effective temperatures and surface gravities:
\begin{itemize}
\item Johnson {\it UBV}: Napiwotzki, Sch\"onberner \& Wenske (1993)
\item Str\"omgren $uvby\beta$: Moon \& Dworetsky (1985) and Napiwotzki et al.\,(1993)
\item Geneva 7-colour: Kobi \& North (1990) and K\"unzli et al.\,(1997)
\end{itemize}
The reddening and absolute magnitudes were estimated by use of the
Str\"omgren $uvby\beta$ system. The calibrations for the Johnson {\it UBV\/} 
and Geneva 7-colour system need an a-priori knowledge of the reddening
which is not easy to estimate. 

An independent way
to derive the interstellar reddening maps is to use data from open clusters as well as 
Galactic field stars.
Several different models have been published in the literature (Arenou, Grenon \& 
G\'omez 1992, Hakkila et al.\,1997). Chen et al.\,(1998) compared the results
from Arenou et al.\,(1992) and those derived from Hipparcos
measurements, and found an overestimation in previously published results
for distances less than 500\,pc. They consequently proposed a new model
for Galactic latitudes of $\pm$10$\degr$, but otherwise find excellent
agreement with the model by Sandage (1972). We use here the
model proposed by Chen et al.\,(1998) who corrected previous models by 
Arenou et al.\,(1992) on the basis of the Hipparcos data.
The values from the calibration of the Str\"omgren
$uvby\beta$ and the model by Chen et al.\,(1998) are in very good
agreement. To minimize possible inconsistencies we have averaged the
values from both approaches.

The Hipparcos parallaxes were converted directly into absolute magnitudes.
The latter were averaged (using weights based on the standard errors) with the absolute
magnitudes derived from the photometric calibrations. 
With the absolute bolometric magnitude of the Sun 
($M_{\rm Bol}$)$_{\odot}$\,=\,4.75\,mag (Cayrel de Strobel 1996) and 
the bolometric correction taken from Drilling \& Landolt (2000), luminosities
(log\,L$_{\ast}$/L$_{\odot}$) were calculated (Table \ref{lb_basic}).

We have also corrected our data for 
the so-called `Lutz--Kelker effect' (Koen 1992), an overestimation 
of parallaxes owing to random errors.
Oudmaijer, Groenewegen \& \hbox{Schrijver} (1998) showed that that bias has to be
taken into account if individual absolute magnitudes from Hipparcos
parallaxes are calculated.

The projected rotational velocities were taken from Uesugi \& Fukuda (1982),
St\"{u}renburg (1993), Abt \& Morrell (1995), Holweger \& Rentzsch-Holm (1995),
Chernyshova et al.\,(1998), Heiter et al.\,(1998), Paunzen et al.\,(1999a,b),
Kamp et al.\,(2001), Solano et al.\,(2001), Andrievsky et al.\,(2002) and Heiter (2002).
If possible the individual values were averaged with weights in accordance with the
listed standard errors.

As the next step, we have used the post-MS evolutionary tracks and isochrones
from Claret (1995) to estimate the individual masses and ages
for our program stars. The models were calculated with solar abundances.
That is justified because the abundance pattern found for the \LB stars
must be restricted to the surface only: almost all of them would lie below
the Population II ZAMS which would be applicable if they were metal-poor
throughout.  Table \ref{lb_calib} lists the masses
together with minimum and maximum ages, which are not necessarily 
equally spaced owing to the shape of the isochrones. 
Since the lifetime for a star on the MS is dependent on its mass, we have
transformed the ages thus
determined into relative ones ($t_{\rm rel}$) in the
following way.  For our sample we find masses 
(Table \ref{lb_calib}) from 1.6 to 2.5\,M$_{\odot}$, which correspond to
times on the MS of 2.2\,Gyr down to 700\,Myr. The relative age is zero for an object 
which just arrived at the ZAMS and unity for stars at the TAMS.
We have taken into account the error
of the estimated mass as well as the error box of the calibrated
ages.

\begin{figure}
\epsfxsize = 84mm
\epsffile{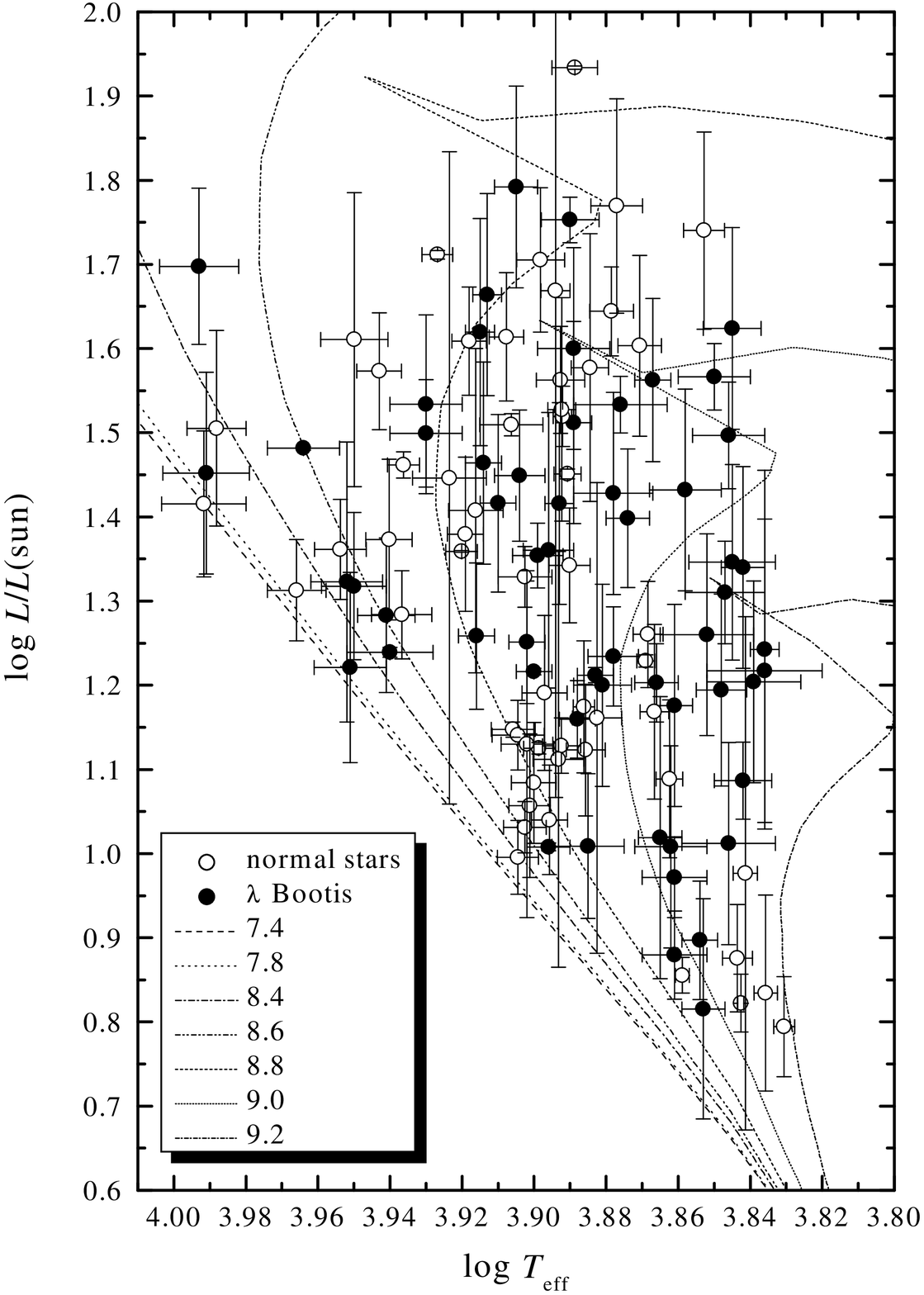}
\caption{The sample of \LB (filled circles) and normal-type 
(open circles) stars within a log\,L$_{\ast}$/L$_{\odot}$
versus log\,$T_{\rm eff}$ diagram. The isochrones (the different 
log\,$t$ values are listed in the legend) were
taken from Claret (1995).} \label{hrd}
\end{figure}

\begin{table}
\caption[]{Calibrated values for the program stars using the models
of Claret (1995); $t_{\rm rel}$ is the relative age
ranging from zero for an object just arriving at the ZAMS to unity
for a star at the TAMS; $\Delta$log\,$g$\,=\,log\,$g$(evol)\,$-$\,log\,$g$(phot);
log\,$t_{1}$ and log\,$t_{2}$ are the minimum and maximum ages derived from the 
error boxes.} \label{lb_calib}
\begin{center}
\begin{tabular}{llccccl}
\hline
\bb HD & \multicolumn{1}{c}{$M$}	&	$\Delta$log\,$g$ & log\,$t$	&	log\,$t_{1}$	&	log\,$t_{2}$	&	\multicolumn{1}{c}{$t_{\rm rel}$}	\\
& \multicolumn{1}{c}{[M$_{\odot}$]} & & [dex] & [dex] & [dex] \\ 
\noalign{\vskip6pt}
\b\bb319&	2.12(9)	&	+0.15	&	8.88	&	8.84	&	8.91	&	0.86(5)	\\
\bb6870	&	1.70(12)	&	+0.22	&	9.02	&	8.70	&	9.05	&	0.69(13)	\\
\bb7908	&	1.61(5)	&	+0.02	&	9.02	&	8.88	&	9.11	&	0.70(8)	\\
\b11413	&	2.02(4)	&	+0.03	&	8.91	&	8.88	&	8.93	&	0.83(3)	\\
\b13755	&	2.21(4)	&	+0.31	&	9.01	&	8.90	&	9.05	&	1.03(8)	\\
\b15165	&	2.14(6)	&	+0.38	&	9.00	&	8.93	&	9.10	&	1.01(9)	\\
\b23392	&	2.30(11)	&	$-$0.09	&	8.00	&	7.00	&	8.50	&	0.35(23)	\\
\b24472	&	1.74(4)	&	+0.10	&	9.13	&	9.09	&	9.17	&	0.88(4)	\\
\b30422	&	1.76(2)	&	+0.21	&	8.69	&	8.52	&	8.80	&	0.54(7)	\\
\b31295	&	2.08(10)	&	+0.00	&	8.51	&	7.60	&	8.71	&	0.47(23)	\\
\b35242	&	1.96(6)	&	+0.19	&	8.81	&	8.69	&	8.86	&	0.70(6)	\\
\b54272	&	1.69(10)	&	+0.16	&	9.12	&	9.00	&	9.20	&	0.83(10)	\\
\b74873	&	2.00(3)	&	+0.00	&	8.52	&	8.20	&	8.70	&	0.51(13)	\\
\b75654	&	1.85(5)	&	+0.15	&	9.03	&	9.01	&	9.05	&	0.85(3)	\\
\b81290	&	1.85(11)	&	$-$0.01	&	9.10	&	9.03	&	9.17	&	0.92(9)	\\
\b83041	&	1.90(11)	&	+0.06	&	9.05	&	8.99	&	9.11	&	0.90(8)	\\
\b83277	&	1.98(12)	&	+0.05	&	9.03	&	8.96	&	9.10	&	0.93(9)	\\
\b84123	&	1.95(5)	&	+0.03	&	9.04	&	9.01	&	9.10	&	0.93(5)	\\
\b87271	&	2.20(3)	&	+0.28	&	8.90	&	8.80	&	9.00	&	0.93(9)	\\
\b90821	&	2.38(16)	&	+0.03	&	8.79	&	8.74	&	8.90	&	0.93(11)	\\
\b91130	&	2.10(10)	&	+0.16	&	8.86	&	8.83	&	8.89	&	0.83(5)	\\
101108	&	2.08(12)	&	$-$0.03	&	8.92	&	8.87	&	8.92	&	0.86(6)	\\
102541	&	1.75(6)	&	$-$0.06	&	8.85	&	8.42	&	8.99	&	0.60(16)	\\
105058	&	2.26(11)	&	$-$0.06	&	8.85	&	8.79	&	9.00	&	0.94(11)	\\
105759	&	2.05(8)	&	+0.15	&	8.96	&	8.92	&	9.10	&	0.94(10)	\\
106223	&	1.85(15)	&	+0.30	&	9.10	&	9.00	&	9.20	&	0.92(12)	\\
107233	&	1.62(4)	&	+0.14	&	8.93	&	8.60	&	9.06	&	0.61(13)	\\
109738	&	1.87(12)	&	+0.09	&	8.98	&	8.90	&	9.02	&	0.81(7)	\\
110377	&	1.84(4)	&	+0.08	&	8.94	&	8.90	&	8.97	&	0.76(3)	\\
110411	&	2.02(10)	&	+0.14	&	8.00	&	7.00	&	8.66	&	0.33(24)	\\
111005	&	1.88(20)	&	+0.05	&	9.09	&	8.99	&	9.18	&	0.93(14)	\\
111604	&	2.42(3)	&	$-$0.02	&	8.79	&	8.77	&	8.90	&	0.96(7)	\\
120500	&	2.30(12)	&	$-$0.06	&	8.81	&	8.75	&	8.90	&	0.90(9)	\\
120896	&	1.83(11)	&	+0.16	&	9.05	&	9.01	&	9.06	&	0.85(6)	\\
125162	&	2.04(7)	&	+0.12	&	8.61	&	8.00	&	8.75	&	0.53(19)	\\
125889	&	1.70(10)	&	+0.19	&	9.03	&	8.99	&	9.10	&	0.77(7)	\\
130767	&	2.28(4)	&	+0.02	&	8.60	&	8.48	&	8.68	&	0.68(7)	\\
142703	&	1.67(2)	&	+0.15	&	9.02	&	8.83	&	9.10	&	0.71(9)	\\
142944	&	2.25(11)	&	+0.31	&	8.99	&	8.90	&	9.06	&	1.05(10)	\\
149130	&	1.97(12)	&	+0.23	&	9.04	&	8.98	&	9.10	&	0.94(8)	\\
153747	&	2.15(12)	&	+0.21	&	8.84	&	8.81	&	8.86	&	0.83(6)	\\
154153	&	1.84(10)	&	+0.30	&	9.08	&	9.03	&	9.13	&	0.89(7)	\\
156954	&	1.56(8)	&	+0.14	&	8.92	&	7.00	&	9.01	&	0.44(29)	\\
168740	&	1.88(2)	&	+0.10	&	8.97	&	8.94	&	8.99	&	0.81(2)	\\
168947	&	2.09(12)	&	+0.13	&	8.94	&	8.88	&	9.01	&	0.91(8)	\\
170680	&	2.50(10)	&	$-$0.08	&	8.48	&	8.30	&	8.57	&	0.66(10)	\\
175445	&	2.25(12)	&	$-$0.03	&	8.78	&	8.73	&	8.82	&	0.83(6)	\\
183324	&	2.09(17)	&	+0.07	&	8.50	&	7.00	&	8.73	&	0.45(29)	\\
184779	&	2.08(12)	&	+0.08	&	8.97	&	8.91	&	9.03	&	0.93(8)	\\
192640	&	1.90(2)	&	+0.10	&	8.90	&	8.85	&	8.91	&	0.75(2)	\\
193256	&	2.18(12)	&	+0.09	&	8.89	&	8.83	&	9.00	&	0.92(10)	\\
193281	&	2.50(12)	&	+0.08	&	8.75	&	8.70	&	8.81	&	0.93(7)	\\
198160	&	2.02(18)	&	$-$0.06	&	8.92	&	8.86	&	8.95	&	0.84(9)	\\
204041	&	1.93(7)	&	+0.06	&	8.90	&	8.82	&	8.92	&	0.76(5)	\\
210111	&	1.90(5)	&	+0.11	&	8.98	&	8.95	&	9.02	&	0.84(4)	\\
216847	&	2.21(10)	&	+0.17	&	8.90	&	8.87	&	9.00	&	0.96(8)	\\
221756	&	2.20(8)	&	+0.06	&	8.78	&	8.73	&	8.82	&	0.80(5)	\\
\hline
\end{tabular}
\end{center}
\end{table}

\begin{table}
\caption[]{Normal-type stars selected as comparison.} \label{standard}
\begin{center}
\begin{tabular}{ll|ll|ll}
\hline
\multicolumn{1}{c}{HD} & \multicolumn{1}{c}{HIP} & 
\multicolumn{1}{c}{HD} & \multicolumn{1}{c}{HIP} &
\multicolumn{1}{c}{HD} & \multicolumn{1}{c}{HIP} \\
\noalign{\vskip6pt}
\b2262	&	\b2072	&	\b71297	&	41375	&	160613	&	\b86565	\\
\b3003	&	\b2578	&	\b79439	&	45493	&	170642	&	\b90887	\\
\b5382	&	\b4366	&	\b85364	&	48341	&	175638	&	\b92946	\\
\b9919	&	\b7535	&	\b87696	&	49593	&	181296	&	\b95261	\\
11636	&	\b8903	&	\b88824	&	50070	&	186689	&	\b97229	\\
13041	&	\b9977	&	\b96113	&	54137	&	187642	&	\b97649	\\
16555	&	12225	&	\b98058	&	55084	&	192425	&	\b99742	\\
17943	&	13421	&	\b98353	&	55266	&	195050	&	100907	\\
19107	&	14293	&	101107	&	56770	&	196078	&	101608	\\
39060	&	27321	&	102124	&	57328	&	201184	&	104365	\\
40136	&	28103	&	105211	&	59072	&	205835	&	106711	\\
45320	&	30666	&	111968	&	62896	&	205852	&	106787	\\
49434	&	32617	&	116706	&	65466	&	205924	&	106856	\\
50241	&	32607	&	118232	&	66234	&	207235	&	107596	\\
50277	&	33024	&	122405	&	68478	&	207958	&	108036	\\
50506	&	31897	&	135379	&	74824	&	210300	&	109412	\\
56405	&	35180	&	135559	&	74689	&	210739	&	109667	\\
59037	&	36393	&	145631	&	79439	&	211356	&	109984	\\
70574	&	41036	&	159561	&	86032	&	220061	&	115250	\\
\hline
\end{tabular}
\end{center}
\end{table}

\begin{figure}
\epsfxsize = 82mm
\epsffile{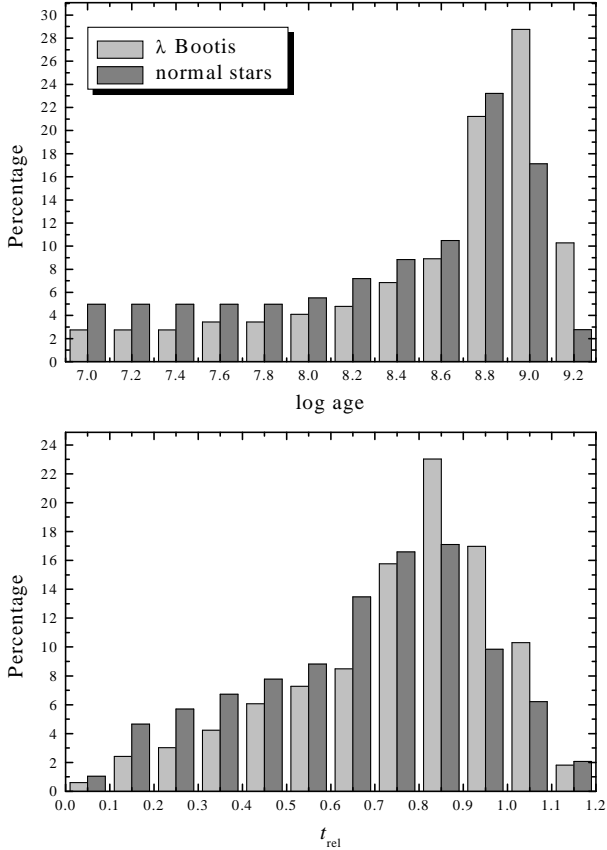}
\caption{Histogram of the calibrated ages (upper panel) and
relative ages for both samples (lower panel).} \label{hist_age}
\end{figure}

\begin{figure}
\epsfxsize = 82mm
\epsffile{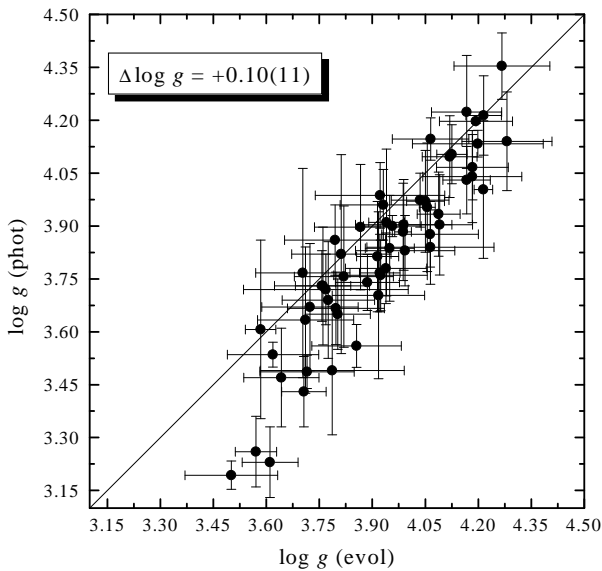}
\caption{Surface gravities calibrated photometrically and calculated via
the mass, luminosity and effective temperature; the mean difference
$\Delta$log\,$g$\,=\,+0.10(11) is not significant.} \label{logg_plot}
\end{figure}

\begin{figure}
\epsfxsize = 82mm
\epsffile{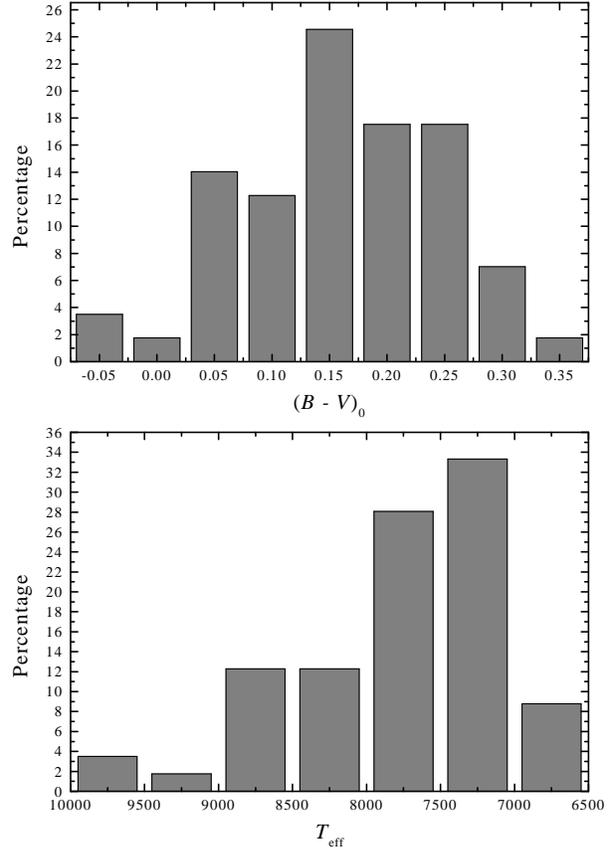}
\caption{Histogram of $(B-V)_0$ values (upper panel) and the calibrated effective 
temperature (lower panel) for the sample of \LB stars.} \label{select}
\end{figure}

\section{A test sample of normal stars}

For comparison with the results obtained for the \LB 
stars we generated a test sample of
apparently normal dwarfs with the same effective
temperatures.  We limited the sample to luminosity class V
objects for which photometric and Hipparcos measurements are
available.  Known spectroscopic binary systems were
excluded.  Then, only objects for which the spectral classifications
published by Gray \& Garrison (1987; 1989a,b) and Garrison \& Gray (1994)
agree within two temperature sub-classes with those by Abt \& Morell
(1995) have been considered.  The final sample (see
Table \ref{standard}) has the same number of stars as the \LB list.
Special attention has been paid to the $(B-V)_0$
distribution of the test sample to ensure that it similar to that of the \LB
objects (Fig. \ref{select}, upper panel). Thus, the test
sample represents luminosity class V stars in the solar
neighbourhood, covering the same spectral range as the
\LB group.

All basic parameters were derived in exactly the same way as for 
the \LB objects. All statistical significance levels presented in the
following sections are based on several hypothesis tests (e.g.~t-test; Rees
1987).

We have not explicitly listed the parameters of this sample but it is
available in electronic form via anonymous ftp 130.79.128.5 or
http://cdsweb.u-strasbg.fr/Abstract.html.  

\section{The Hertzsprung--Russell diagram based on Hipparcos
and photometric data}

Earlier investigations already concluded that the group of \LB stars
comprises the whole area between the ZAMS and TAMS. Only the interpretation
of the evolutionary stage (PMS or post-MS) varied. In this work we have
not investigated a possible PMS hypothesis as proposed by Paunzen (1997).
Beside the arguments given in Bohlender et al.~(1999) and Faraggiana \& Bonifacio (1999)
we would like to add a few more points. If we speculate that some of these stars
are in a PMS phase then they would have ages of one Myr or even younger
since they are in the `right upper corner' of the Hertzsprung--Russell diagram
(HRD hereafter) very close to the birth-line according to the
models of Palla \& Stahler (1993). So they should be very 
bright in the IR region and lie within the boundaries of star-forming 
regions or molecular clouds; neither implication has been verified
so far (King 1994).

Figure \ref{hrd} shows the HRD with the data taken from Table~\ref{lb_basic}.
The post-MS evolutionary isochrones are from Claret (1995). 
It is encouraging to see that no
object lies below the ZAMS; that gives further confidence in the 
calibrations used.

To estimate the most probable age distribution of our program stars, we 
have used a moving average (Fig. \ref{hist_age}). 
Such a method takes into account the errors of individual data as it
counts all points (plus and minus the standard deviation) which lie in 
a certain bin (Rees 1987).
From the histogram it is clear that the group of \LB stars comprises evolutionary
stages of the entire MS with a peak at $t_{\rm rel}$\,$\approx$\,0.85 (about 1\,Gyr).
Taking the few candidates in the Orion OB1 association and NGC~2264, the percentage
for very young objects (log\,$t$\,$<$\,7.0) would be approximately 5\% which fits 
very well into the overall picture. 

Since the distribution of the test sample (Fig. \ref{hist_age}) agrees at a 99.9\%
significance level, the age distribution of the \LB objects is {\it not} 
distinct from that of other luminosity class V objects in the same spectral domain
within the solar neighbourhood. That is already a first hint that the mechanism behind the
\LB phenomenon operates only within very tight constraints, since only 2\% of all stars
are objects of the \LB type.

Gray \& Corbally (1998, 1999) made an extensive survey for \LB stars in 24
open clusters and associations and found not a single candidate. The cluster ages
range from 15 to 700\,Myr. Therefore, only a few candidates in 
the Orion OB1 association and NGC~2264 remain.
It seems that the star-formation process within
open clusters does not favour the manifestation of the \LB phenomenon. 

\begin{table*}
\caption[]{Comparison of the calibrated stellar parameters from this
work (columns TW), Iliev \& Barzova (1995; columns IL95) and Paunzen
(1997; columns PA97).} \label{lb_comparison}
\begin{center}
{\tiny
\begin{tabular}{llcclcllccccclc}
\hline
\bb HD & \multicolumn{3}{c}{log\,$T_{\rm eff}$}		& \multicolumn{3}{c}{log\,L$_{\ast}$/L$_{\sun}$}	&	
\multicolumn{3}{c}{$M$}	&	\multicolumn{3}{c}{log\,$t$}	& \multicolumn{2}{c}{$t_{\rm rel}$} \\
& \multicolumn{1}{c}{TW} & IL95 & PA97 & \multicolumn{1}{c}{TW} & IL95 & 
\multicolumn{1}{c}{PA97} & \multicolumn{1}{c}{TW} & IL95 & PA97 
& TW & IL95 & PA97 & \multicolumn{1}{c}{TW} & IL95\\
& & $\pm$0.02 & $\pm$0.02 & & $\pm$0.01 & & & $\pm$0.2 & $\pm$0.1 & & $\pm$0.05 & $\pm$0.05 & \\
\noalign{\vskip3pt}
\b\bb319&3.904(7)	& 3.92 & 3.91 &	1.45(8)	&	1.53 & 1.34(5) & 2.12(9)	
&	2.2 & 2.0 & 8.88	&	8.81 & 8.76 & 0.86(5)	& 0.77 \\
\b11413	& 3.899(4)	& 3.91 & 3.91 &	1.35(4)	&	1.41 & 1.28(4) & 2.02(4)	
&	2.1 & 1.9 & 8.91	&	8.87 & 8.82 & 0.83(3)	& 0.71 \\
\b30422	& 3.896(6)	& 3.91 & 3.90 &	1.01(1)	&	1.03 & 0.96(3) & 1.76(2)	
&	1.8 & 1.8 & 8.69	&	8.58 & 7.00 & 0.54(7)	& 0.23 \\
\b31295	& 3.950(9)	& 3.96 & 3.95 &	1.32(9)	&	1.40 & 1.20(3) & 2.08(10)	
&	2.2 & 2.0 & 8.51	&	8.54 & 7.00 & 0.47(23)	& 0.38 \\
107233	& 3.861(9)	& 3.86 & 3.87 &	0.88(5)	&	0.94 & 0.80(5) & 1.62(4)	
&	1.6 & 1.6 & 8.93	&	9.04 & 7.00 & 0.61(13)	& 0.52 \\
110411 	& 3.951(10)	& 3.96 & 3.94 &	1.22(11)	&	1.30 & 1.09(3) & 2.02(10)	
&	2.1 & 1.9 & 8.00	&	8.30 & 7.00 & 0.33(24)	& 0.19 \\
125162	&	3.935(5)	& 3.95 & 3.94 &	1.28(9)	&	1.38 & 1.20(1) & 2.04(7)	
&	2.1 & 2.0 & 8.61	&	8.58 & 7.20 & 0.53(19)	& 0.40 \\
142703	&	3.868(9)	& 3.87 & 3.86 &	0.97(5)	&	1.00 & 0.93(3) & 1.67(2)	
&	1.7 & 1.7 & 9.02	&	9.02 & 8.50 & 0.71(9)	& 0.56 \\
183324	&	3.950(5)	& 3.97 & 3.96 &	1.32(17)	&	1.51 & 1.20(4) & 2.09(17)	
&	2.3 & 2.1 & 8.50	&	8.52 & 7.00 & 0.45(29)	& 0.42 \\
192640	&	3.903(2)	& 3.91 & 3.90 &	1.22(1)	&	1.23 & 1.16(2) & 1.90(2)	
&	1.9 & 1.9 & 8.90	&	8.85 & 8.65 & 0.75(2)	& 0.54 \\
193256	&	3.894(9)	& 3.92 & 3.90 &	1.51(12)	&	1.48 & 1.50(27) & 2.18(12)	
&	2.1 & 2.2 & 8.89	&	8.86 & 8.79 & 0.92(10) & 0.78 \\
193281	&	3.907(8)	& 3.92 & 3.91 &	1.79(12)	&	1.81 & 1.96(27) & 2.50(12)	
&	2.5 & 2.5 & 8.75	&	8.72 & 8.84 & 0.93(7)	& 0.91 \\
198160	&	3.907(13)	& 3.90 & 3.90 &	1.36(16)	&	1.24 & 1.37(7) & 2.02(18)	
&	1.9 & 2.0 & 8.92	&	8.86 & 8.77 & 0.84(9)	& 0.58 \\
204041	&	3.906(6)	& 3.91 & 3.91 &	1.25(7)	&	1.21 & 1.20(7) & 1.93(7)	
&	1.9 & 1.9 & 8.90	&	8.81 & 8.57 & 0.76(5)	& 0.48 \\
210111	&	3.888(3)	& 3.89 & 3.88 &	1.23(6)	&	1.30 & 1.16(5) & 1.90(5)	
&	2.0 & 1.8 & 8.98	&	8.94 & 8.84 & 0.84(4)	& 0.69 \\
221756	&	3.931(9)	& 3.96 & 3.96 &	1.50(6)	&	1.67 & 1.41(4) & 2.20(8)	
&	2.4 & 2.2 & 8.78	&	8.65 & 8.10 & 0.80(5)	& 0.68 \\
\hline
\end{tabular}
}
\end{center}
\end{table*}

Table \ref{lb_comparison} lists the 16 stars that we have in common with the papers of
Iliev \& Barzova (1995) and Paunzen (1997). For the calibration of effective
temperatures they both used the method of Moon \& Dworetsky (1985) within
the Str\"omgren $uvby\beta$ photometric system; Iliev \& Barzova (1995) also derived the 
luminosities with this calibration whereas Paunzen (1997) took advantage of the Hipparcos
data. For the calibration of mass and
age, Iliev \& Barzova (1995) interpolated between the evolutionary tracks given by 
Schaller et al.~(1992), whereas Paunzen (1997) used the CESAM models by Morel (1997).
As expected, there are star-to-star variations, but overall the parameters
fit well.
The effective temperatures from this work are in very good agreement with those of both
references. The luminosities and thus the calibrated ages are in better agreement
with those of Iliev \& Barzova (1995). 

As a further test of the calibrated values, we have calculated the surface
gravity via the effective temperature, mass and luminosity. Figure~\ref{logg_plot}
shows the correlation of the photometrically calibrated surface gravity
log\,$g$(phot) and the calculated one log\,$g$(evol). The mean errors for both
parameters are typically $\pm$0.1\,dex.
Although there is some indication of a systematic offset, the mean difference
$\Delta$log\,$g$\,=\,log\,$g$(evol)\,$-$\,log\,$g$(phot)\,=\,+0.10(11) is not
significant. A similar effect was already noticed by Iliev \& Barzova
(1995) whereas Faraggiana \& Bonifacio (1999) reported an inconsistency between
the position in the HRD and the log\,$g$ values derived from Str\"omgren $uvby\beta$
photometry using the calibration of Moon \& Dworetsky (1985). We conclude that the
photometric calibrations for the group of \LB stars are valid, in contradiction
to the results of Faraggiana \& Bonifacio (1999).

Before making a detailed statistical analysis, we looked to see whether our
sample exhibits an apparent bias. Figure \ref{select} shows that the sample
includes significantly more cooler objects (70\%) with effective
temperatures lower than 8000\,K.  
However, there is no observational bias from classification-resolution spectroscopy
since the spectroscopic survey for new members included many more hotter- than 
cooler-type objects (Gray \& Corbally 1999, Paunzen 2001). 
That fact could be interpreted as a manifestation of the working mechanism 
itself, or it could be due to
the method used in this very limited spectral range. It is well known
that at cooler temperatures (spectral domain A5 to F2), even a moderate metal-weakness
can be detected at classification resolution since the overall metallic-line
spectrum is much richer than for an A0-type object. However,
we are not able to decide whether the distribution of the \LB sample is due
to a bias within the observational technique or due to the phenomenon itself.

\begin{figure}
\epsfxsize = 82mm
\epsffile{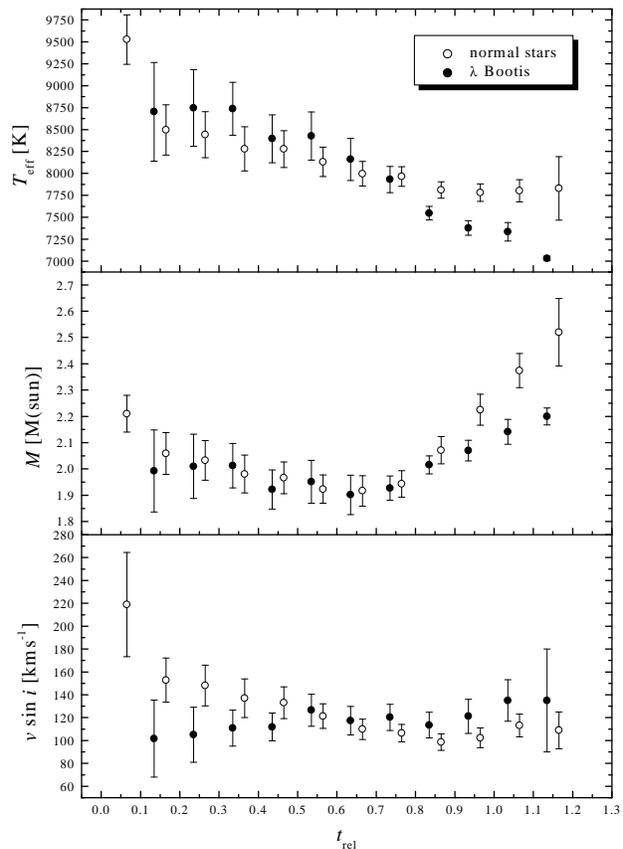}
\caption{Correlation found for the effective temperature (lower panel)
and mass (middle panel) with the relative age; no correlation is
detected with the projected rotational velocity (upper panel).
The \LB (filled circles) and normal-type
(open circles) stars are binned with respect to their relative
age.} \label{corr_plot}
\end{figure}

Figure \ref{corr_plot} shows the averaged effective temperatures, masses and
projected rotational velocities for each age bin. The mean effective temperature
is rather constant up to $t_{\rm rel}$\,$\approx$\,0.5 with values between 8300\,K to
8800\,K. It then decreases linearly almost to 7000\,K for the most evolved
objects. 
The mean masses are constant within the error bars at about 1.9\,M$_{\odot}$
for $t_{\rm rel}$\,$<$\,0.75 and then increase to almost 2.2\,M$_{\odot}$. Most
interesting is the non-existence of a correlation of the projected rotational velocity 
with age. The mean value for the whole range is about 120\,km\,s$^{-1}$. 
If we compare these results with those of our test sample, we find several differences
as well as similarities:
\begin{itemize}
\item The trends for the effective temperature and mass are identical. However,
the \LB objects seem to have lower temperatures as well as masses for
$t_{\rm rel}$\,$>$\,0.8
\item The $v$\,sin\,$i$ distributions are identical within 1$\sigma$ with a slightly
higher scatter for the \LB group ($\overline{\sigma}_{{\it LB}}$\,=\,19\,km\,s$^{-1}$ and
$\overline{\sigma}_{{\it Nor}}$\,=\,15\,km\,s$^{-1}$)
\end{itemize} 
Overall, there is no obvious distinction between the two samples.

Gray (1988) reported that several \LB stars exhibit peculiar hydrogen-line profiles
with weak cores and broad but often shallow wings. Iliev \& Barzova (1993)
examined the
peculiar profiles of four objects. They were able to fit those profiles with two
models having
different temperatures (hotter for the wings by approximately 400\,K) and
concluded that it is a sign of circumstellar material around the objects. Faraggiana
\& Bonifacio (1999) interpret the peculiar profiles as an indication of undetected
spectroscopic-binary systems in which two stars with solar abundances but different stellar
parameters mimic one apparently metal-weak object. The classification of the hydrogen-line
profiles for 19 stars
were taken from Gray \& Corbally (1993) and Paunzen \& Gray (1997). For our
sample we find only two objects (HD~30422 and HD~107233) with peculiar profiles 
among the younger objects ($t_{\rm rel}$\,$<$\,0.66) but seven with normal ones. 
The picture for the 
older objects is just the opposite: there are only two objects (HD~90821 and HD~120500) 
with normal profiles but nine with peculiar ones.

Several attempts were made to detect signs of circumstellar lines in the
optical domain (Andrillat, Jaschek \& Jaschek 1995; Hauck, Ballereau \& Chauville 1995, 
1998; Holweger, Hempel \& Kamp
1999). From our list, sixteen objects were investigated but
only three objects have positive detections: HD~11413, HD~193256
and HD~198160 (Holweger et al.~1999). These stars are rather evolved 
($t_{\rm rel}$\,=\,0.83, 0.92 and 0.84, respectively) but owing to the poor number
statistics, any conclusion about the significance has to be treated with caution.

\begin{table*}
\caption[]{Abundances from the literature for our program
stars. Different values for a single
element were added and a mean was calculated. All values are given 
with respect to the Sun as: [X]\,=\,log\,X\,--\,log\,X$_{\odot}$.
In parentheses are the estimated errors of the means or the individual
error taken from the relevant reference.} \label{lb_abund}
\begin{center}
\begin{tabular}{rccccccccc}
\hline
HD\b	&	C	&	N	&	O	&	Na	&	Mg	&	Al	&	Si	&	S	&	Ca	\\
\noalign{\vskip3pt}
319\b	&	$-$0.16(15)	&	--	&	+0.22(10)	&	+1.00(10)	&	$-$0.73\b(5)	&	$-$0.58(20)	&	$-$0.45(10)	&	--	&	$-$0.68(20)	\\
11413	&	$-$0.15(15)	&	--	&	$-$0.10(10)	&	--	&	$-$1.18(20)	&	--	&	$-$0.20(20)	&	--	&	$-$0.74(30)	\\
15165	&	+0.00(30)	&	--	&	$-$0.30(30)	&	+0.00(30)	&	$-$1.50(30)	&	--	&	$-$0.90(30)	&	+0.00(30)	&	$-$1.00(30)	\\
31295	&	$-$0.17(15)	&	$-$0.24(15)	&	$-$0.29(15)	&	$-$0.52(20)	&	$-$1.15(20)	&	--	&	$-$0.95(20)	&	$-$0.16(20)	&	$-$0.90(20)	\\
74873	&	+0.60(10)	&	--	&	$-$0.10(30)	&	+0.50(30)	&	$-$0.90(10)	&	$-$1.10(30)	&	--	&	--	&	$-$1.00(30)	\\
75654	&	$-$0.44(10)	&	+0.00(15)	&	+0.14(10)	&	--	&	$-$0.52(18)	&	--	&	--	&	$-$0.10(15)	&	$-$0.84(15)	\\
84123	&	$-$0.01(24)	&	+0.00(30)	&	$-$0.30(20)	&	$-$0.55(40)	&	$-$0.76(20)	&	--	&	$-$0.75(20)	&	$-$0.50(10)	&	$-$0.58(20)	\\
91130	&	+0.20(30)	&	+0.30(15)	&	+0.11(30)	&	--	&	$-$0.63\b(6)	&	--	&	$-$0.45\b(7)	&	+0.18(15)	&	$-$3.00(15)	\\
101108	&	+0.10(10)	&	--	&	--	&	--	&	$-$0.30(20)	&	$-$0.90(30)	&	$-$0.50(30)	&	--	&	$-$0.60(10)	\\
106223	&	+0.30(10)	&	--	&	--	&	--	&	$-$1.40(30)	&	$-$2.10(30)	&	$-$1.70(30)	&	--	&	$-$1.59(10)	\\
107233	&	$-$0.30(30)	&	--	&	--	&	--	&	$-$0.80(20)	&	$-$2.20(30)	&	--	&	--	&	$-$1.40(10)	\\
110411	&	+0.10(20)	&	+0.30(30)	&	--	&	--	&	$-$0.50(10)	&	--	&	$-$0.30(30)	&	--	&	$-$0.84(20)	\\
111604	&	  $-$0.25\b(5)	&	--	&	--	&	+0.47(15)	&	$-$0.98\b(5)	&	--	&	$-$0.85(30)	&	--	&	--	\\
120500	&	+0.20(20)	&	+0.00(15)	&	$-$0.09(15)	&	+0.32(30)	&	$-$0.23\b(7)	&	--	&	$-$0.95\b(5)	&	$-$0.13(15)	&	$-$0.30(15)	\\
125162	&	$-$0.34(20)	&	$-$0.12(15)	&	$-$0.22(20)	&	$-$1.30(15)	&	$-$2.00(20)	&	--	&	$-$1.00(20)	&	$-$0.45(15)	&	$-$1.98(15)	\\
142703	&	$-$0.28(20)	&	$-$0.50(15)	&	$-$0.19(10)	&	--	&	$-$1.20(20)	&	--	&	--	&	$-$0.53(15)	&	$-$1.20(10)	\\
168740	&	$-$0.42(10)	&	--	&	$-$0.03(10)	&	--	&	$-$0.90(10)	&	--	&	--	&	--	&	$-$0.60(20)	\\
170680	&	$-$0.06(10)	&	$-$0.07(10)	&	--	&	--	&	$-$0.20(20)	&	--	&	--	&	--	&	--	\\
183324	&	+0.11(20)	&	$-$0.03(30)	&	$-$0.10(30)	&	+0.10(30)	&	$-$1.73(20)	&	$-$1.50(30)	&	$-$1.10(20)	&	$-$0.13(15)	&	$-$1.32(30)	\\
192640	&	$-$0.19(20)	&	$-$0.22(15)	&	$-$0.32(30)	&	$-$1.10(20)	&	$-$1.62(20)	&	--	&	$-$1.00(30)	&	$-$0.31(15)	&	$-$1.35(20)	\\
193256	&	$-$0.45(20)	&	--	&	$-$0.23(10)	&	+0.90(30)	&	$-$0.15(20)	&	--	&	+0.00(30)	&	--	&	$-$0.51(30)	\\
193281	&	$-$0.49(20)	&	+0.30(15)	&	$-$0.05(10)	&	+1.00(20)	&	$-$0.08(10)	&	--	&	--	&	+0.14(15)	&	$-$0.68(20)	\\
198160	&	$-$0.18(30)	&	--	&	$-$0.18(10)	&	+0.30(20)	&	$-$0.13(10)	&	--	&	$-$0.20(20)	&	--	&	$-$0.67(30)	\\
204041	&	$-$0.58(20)	&	$-$0.05(15)	&	$-$0.42(20)	&	+0.36(20)	&	$-$1.04(20)	&	$-$0.93(20)	&	$-$0.63(20)	&	+0.07(20)	&	$-$0.98(20)	\\
210111	&	$-$0.23(10)	&	--	&	$-$0.20(10)	&	+0.60(10)	&	$-$1.00(12)	&	$-$0.52(20)	&	$-$0.65(10)	&	--	&	$-$0.96(20)	\\
221756	&	+0.00(20)	&	+0.20(15)	&	+0.17(20)	&	+1.17(20)	&	$-$0.70(20)	&	--	&	$-$1.00(20)	&	+0.06(15)	&	$-$0.49(20)	\\
\hline
\noalign{\vskip3pt}
	&	Sc	&	Ti	&	Mn	&	Cr	&	Fe	&	Ni	&	Sr	&	Y	&	Ba	\\
\noalign{\vskip3pt}
319	&	$-$1.10(20)	&	$-$0.75(20)	&	--	&	$-$0.40(20)	&	$-$0.64(20)	&	--	&	$-$0.35(30)	&	--	&	$-$0.25(20)	\\
11413	&	$-$1.10(30)	&	$-$1.40(20)	&	--	&	$-$1.20(20)	&	$-$1.43(20)	&	--	&	$-$1.50(20)	&	--	&	$-$0.90(20)	\\
15165	&	$-$1.40(30)	&	$-$0.50(30)	&	--	&	$-$0.90(30)	&	$-$1.60(30)	&	$-$1.00(30)	&	$-$0.80(30)	&	--	&	--	\\
31295	&	$-$0.90(30)	&	$-$0.96(20)	&	--	&	$-$0.75(20)	&	$-$1.07(20)	&	$-$0.50(20)	&	$-$1.51(20)	&	--	&	$-$0.45(20)	\\
74873	&	$-$0.40(30)	&	$-$0.90(10)	&	--	&	$-$0.30(20)	&	$-$0.70(10)	&	$-$0.20(30)	&	$-$0.90(40)	&	--	&	--	\\
75654	&	--	&	$-$1.01(29)	&	$-$1.00(42)	&	$-$1.10(30)	&	$-$1.08(14)	&	--	&	--	&	--	&	--	\\
84123	&	$-$0.90(31)	&	$-$0.86(10)	&	$-$0.84(20)	&	$-$0.85(30)	&	$-$0.72(20)	&	$-$0.74(30)	&	$-$0.50(30)	&	$-$0.55(13)	&	$-$0.60(20)	\\
91130	&	--	&	$-$0.64\b(8)	&	--	&	$-$0.48\b(3)	&	$-$1.16\b(8)	&	--	&	--	&	--	&	--	\\
101108	&	$-$0.60(40)	&	$-$0.20(20)	&	$-$0.40(20)	&	$-$0.60(10)	&	$-$0.70(10)	&	$-$0.30(10)	&	$-$0.70(40)	&	+0.00(40)	&	$-$0.70(30)	\\
106223	&	$-$1.10(30)	&	$-$1.02(10)	&	$-$1.70(10)	&	$-$1.34(20)	&	$-$1.60(10)	&	$-$0.97(20)	&	$-$1.70(30)	&	$-$0.50(30)	&	$-$1.60(20)	\\
107233	&	$-$1.70(30)	&	$-$1.34(30)	&	$-$1.30(20)	&	$-$1.28(30)	&	$-$1.38(21)	&	--	&	$-$1.30(30)	&	$-$0.90(30)	&	--	\\
110411	&	$-$1.10(30)	&	$-$0.90(10)	&	--	&	--	&	$-$1.07(20)	&	--	&	$-$1.10(20)	&	--	&	--	\\
111604	&	--	&	--	&	--	&	$-$1.07(30)	&	$-$1.08\b(4)	&	--	&	--	&	--	&	--	\\
120500	&	--	&	--	&	--	&	--	&	$-$0.67\b(6)	&	--	&	--	&	--	&	--	\\
125162	&	$-$0.70(20)	&	$-$2.01(20)	&	--	&	$-$1.00(20)	&	$-$1.98(20)	&	$-$1.20(20)	&	$-$2.10(15)	&	--	&	$-$0.90(20)	\\	
142703	&	$-$1.50(30)	&	$-$1.27(10)	&	--	&	$-$1.40(10)	&	$-$1.32(12)	&	$-$0.90(20)	&	$-$1.20(30)	&	--	&	$-$1.50(30)	\\
168740	&	$-$1.10(30)	&	$-$0.88(10)	&	--	&	$-$1.10(10)	&	$-$0.88(10)	&	--	&	$-$1.00(30)	&	--	&	$-$0.50(30)	\\
170680	&	--	&	$-$0.40(10)	&	--	&	$-$0.40(30)	&	$-$0.50(10)	&	--	&	--	&	--	&	--	\\
183324	&	$-$1.46(40)	&	$-$1.42(20)	&	--	&	$-$1.38(20)	&	$-$1.50(30)	&	--	&	$-$2.00(30)	&	--	&	$-$0.67(20)	\\
192640	&	$-$1.15(20)	&	$-$1.33(20)	&	$-$2.00(20)	&	$-$1.65(20)	&	$-$1.63(20)	&	$-$0.60(20)	&	$-$1.49(30)	&	--	&	$-$1.03(20)	\\
193256	&	$-$0.90(40)	&	$-$0.30(30)	&	--	&	$-$1.00(30)	&	$-$0.90(30)	&	--	&	$-$0.50(30)	&	--	&	$-$0.88(50)	\\
193281	&	--	&	$-$0.36(20)	&	--	&	$-$0.20(20)	&	$-$0.93(20)	&	--	&	$-$0.10(20)	&	--	&	$-$0.54(20)	\\
198160	&	--	&	$-$0.50(20)	&	--	&	$-$0.90(20)	&	$-$0.78(30)	&	--	&	$-$1.30(30)	&	--	&	$-$0.98(30)	\\
204041	&	$-$1.03(40)	&	$-$1.21(20)	&	--	&	$-$0.72(30)	&	$-$0.87(20)	&	$-$0.40(20)	&	--	&	--	&	$-$0.09(20)	\\
210111	&	$-$1.30(20)	&	$-$1.05(20)	&	--	&	$-$1.18(20)	&	$-$0.89(20)	&	--	&	+0.45(20)	&	--	&	$-$0.20(20)	\\
221756	&	$-$0.60(20)	&	$-$0.50(20)	&	--	&	$-$0.50(20)	&	$-$0.59(30)	&	--	&	--	&	--	&	$-$0.15(30)	\\
\hline
\end{tabular}
\end{center}
\end{table*}

\begin{table*}
\caption[]{Kinematic data from the Hipparcos database
for the program stars. In
parentheses are the errors in the final digits of the corresponding quantity.} \label{lb_mot}
\begin{center}
\begin{tabular}{rrcrrcccc}
\hline
HD &  \multicolumn{1}{c}{$\pi$} & $d$ & \multicolumn{1}{c}{$\mu_{\alpha}$}	&	
\multicolumn{1}{c}{$\mu_{\delta}$}	&	
\multicolumn{1}{c}{$RV$}	&	\multicolumn{1}{c}{$U$}	&	\multicolumn{1}{c}{$V$}	&	
\multicolumn{1}{c}{$W$} \\
&  \multicolumn{1}{c}{[mas]} & [pc] & \multicolumn{1}{c}{[mas\,yr$^{-1}$]} & 
\multicolumn{1}{c}{[mas\,yr$^{-1}$]} & \multicolumn{1}{c}{[km\,s$^{-1}$]} & 
\multicolumn{1}{c}{[km\,s$^{-1}$]}
& \multicolumn{1}{c}{[km\,s$^{-1}$]} & \multicolumn{1}{c}{[km\,s$^{-1}$]} \\
\hline
319	
&	12.45(0.74)	&	80(5) & +70.01(0.94)	&	$-$34.71(0.41)	&	$-$10.7(9.9)&	$-$18.1(1.8)	&	$-$25.5(1.8)	&	\b+4.9(9.7) \\
6870		
&	10.30(0.61)	&	97(6) & +120.99(0.51)	&	$-$12.65(0.59)	&	+12.4(3.0)	&	$-$39.2(2.6)	&	$-$41.7(2.8)	&	\b$-$0.3(2.3)	\\
7908	
&	11.41(0.98)	&	88(8) & $-$5.54(1.04)	&	$-$37.51(0.50)	&	+13.2(3.0)	&	\b+9.2(1.0)	&	$-$11.3(1.1)	&	$-$14.5(3.0)	\\
11413	
&	13.37(0.64)	&	75(4) & $-$47.97(0.51)	&	$-$4.30(0.57)	&	\b+7.2(5.6)	&	+14.3(0.8)	&	\b+6.5(2.4)	&	\b$-$9.8(5.1)	\\
15165	
&	8.51(0.89)	&	118(12) & +35.50(1.10)	&	$-$11.70(0.74)	&	+33.5(2.1)	&	$-$32.4(1.8)	&	\b$-$8.6(1.4)	&	$-$20.8(2.1) \\
30422	
&	17.40(0.68)	&	57(2) & $-$4.04(0.48)	&	+18.19(0.60)	&	+16.5(3.5)	&	$-$12.4(1.8)	&	\b$-$6.4(2.1)	&	$-$10.2(2.2)	\\
31295	
&	27.04(0.94)	&	37(1) & +40.08(1.04)	&	$-$128.37(0.65)	&	+12.9(1.7)	&	\b$-$6.0(1.6)	&	$-$23.9(0.7)	&	$-$10.8(0.8)	\\
35242	
&	13.32(0.98)	&	75(6) & $-$30.42(0.99)	&	+16.63(0.63)	&	  \b+9.0(3.0)	&	\b$-$9.7(2.7)	&	\b+8.0(1.1)	& \b$-$8.7(1.2)	\\
74873	
&	16.38(1.16)	&	61(4) & $-$65.20(1.25)	&	$-$51.06(0.70)	&	+23.3(3.0)	&	$-$23.0(2.4)	&	$-$22.9(1.4)	& \b$-$7.9(2.1)	\\
75654	
&	12.82(0.58)	&	78(4) & $-$63.91(0.42)	&	+41.13(0.49)	&	\b+9.4(1.1)&	$-$28.0(0.8)	&	\b$-$5.3(1.2)	&	\b$-$8.2(0.9)	\\
84123	
&	9.09(0.90)	&	110(11) & $-$20.74(1.11)	&	$-$86.15(0.43)	&	+16.3(3.0)	&	$-$16.0(2.3)	&	$-$45.5(4.4)	&	\b+8.4(2.3) \\
87271 
& 6.80(0.88) & 147(19) & +3.33(0.86) & +5.01(0.49) & \b+0.1(3.0) & \b+0.4(1.5) & \b+3.3(1.5) & \b+2.5(2.3) \\
91130	
&	13.33(0.76)	&	75(4) & +17.10(0.60)	&	+6.89(0.42)
&	$-$14.5(3.0)	&	+11.6(1.5)	&\b +5.6(0.4)	&	\b$-$9.4(2.6)	\\
105058	
&	\b5.32(1.04)	&	188(37) & $-$9.80(0.84)	&	+0.66(0.81)	&	\b$-$7.2(3.0)	&	\b$-$5.5(1.9)	&	\b$-$5.3(1.3)	&	\b$-$8.3(2.8)	\\
106223	
&	9.10(0.86)	&	110(10) & $-$67.32(1.04)	&	+19.29(0.55)	&	$-$15.4(3.7)	&	$-$32.8(2.9)	&	\b$-$7.3(1.9)	&	$-$20.9(3.7)	\\
110377	
&	14.60(0.80)	&	68(4) & $-$106.08(0.78)	&	+0.71(0.48)	&	\b+3.0(8.5)	&	$-$29.0(1.9)	&	$-$18.7(2.5)	&	\b+1.6(8.1) \\
110411	
&	27.10(0.82)	&	37(1) & +82.62(0.80)	&	$-$89.51(0.48)	&	  \b$-$12.7(11.5)	&	+18.7(1.5)	&	$-\b$1.6(3.1)	&	\b$-$16.2(11.0)	\\
111005	
&	5.75(1.01)	&	174(31) & $-$43.12(0.86)	&	$-$6.92(0.58)	&	\b+0.6(3.0)	&	$-$27.1(5.3)	&	$-$23.6(3.1)	&	\b$-$2.6(3.5)	\\
111604	
&	8.43(0.73)	&	119(10) & $-$88.59(0.67)	&	+21.83(0.56)	&	$-$17.1(4.7)	&	$-$46.7(3.7)	&	$-$19.4(2.4)	&	$-$19.3(4.8)	\\
120500	
&	6.97(0.90)	&	143(19) & $-$26.63(0.70)	&	+10.97(0.68)	&	\b+7.1(3.3)	&	$-$15.0(2.4)	&	\b$-$6.4(1.9)	&	+13.0(2.9)	\\
120896	
&	4.87(1.15)	&	205(48) & $-$18.65(1.14)	&	+3.00(0.78)	&	$-$24.4(3.0)	&	$-$25.0(3.7)	&	\b$-$5.5(3.0)	&	$-$16.7(2.6)	\\
125162	
&	33.58(0.61)	&	30(1) & $-$187.42(0.52)	&	+159.01(0.43)	&	\b$-$0.3(4.0)	&	$-$34.6(0.5)	&	\b$-$5.1(1.7)	&	\b$-$2.6(3.6)	\\
130767	
& 7.82(0.78) & 128(13) & $-$46.63(0.69) & $-$6.87(0.62) & $-$14.0(3.0) & $-$20.5(2.0) & $-$24.4(2.1) & \b$-$0.7(3.0) \\
142703	
&	18.89(0.78)	&	53(2) & +71.62(0.83)	&	$-$305.20(0.67)	&	+15.9(5.4)	&	+23.5(4.8)	&	\b+4.0(0.7)	&	\b$-$9.2(2.6)	\\
153747	
&	5.32(0.98)	&	188(35) & +6.55(0.99)	&	+14.08(0.68)	&	\b$-$6.3(0.5)	&	\b$-$3.2(1.1)	&	+14.6(1.8)	&	\b+2.8(1.9)	\\
154153	
&	15.16(1.11)	&	66(5) & +11.40(0.90)	&	+25.04(0.61)	&	$-$32.6(3.0)	&	$-$28.6(2.9)	&	+17.5(1.0)	&	\b+3.0(0.5)	\\
168740	
&	14.03(0.69)	&	71(4) & +0.57(0.59)	&	$-$101.87(0.49)	&	$-$21.1(3.0)	&	$-$36.3(2.6)	&	$-$17.5(1.9)	&	\b$-$2.8(1.2)	\\
170680	
&	15.30(0.81)	&	65(3) & $-$6.98(0.84)	&	$-$21.43(0.64)	&	$-$34.6(3.5)	&	$-$31.8(3.4)	&	$-$15.2(0.9)	&	\b+1.3(0.4)	\\
183324	
&	16.95(0.87)	&	59(3) & $-$0.51(0.72)	&	$-$33.35(0.43)	&	+12.0(3.0)	&	+14.1(2.3)	&	\b+0.7(1.9)	&	\b$-$5.7(0.5)	 \\
192640	
&	24.37(0.55)	&	41(1) & +69.22(0.46)	&	+68.67(0.45)	&	$-$18.1(1.1)	&	$-$22.7(0.4)	&	$-$12.4(1.1)	&	\b$-$4.2(0.3)	\\
193281	
&	4.58(1.59)	&	218(76) & $-$4.32(1.54)	&	$-$0.75(1.07)	&	\b+2.0(2.7)	&	\b+4.2(2.5)	&	$\b-$1.0(1.3)	&	\b+2.5(2.3)	\\
198160	
&	13.67(1.16)	&	73(6) & +83.04(0.86)	&	$-$49.38(0.92)	&	$-$16.0(3.0)	&	$-$34.9(2.4)	&	\b$-$8.6(2.3)	&	\b$-$9.3(2.5)	\\
204041	
&	11.46(0.99)	&	87(8) & +50.00(1.18)	&	+13.84(0.77)	&	  \b$-$15.1(26.7)	&	\b$-$25.1(13.3)	&	\bb$-$6.7(17.9)	&	\bb$-$3.4(14.8)	  \\
210111	
&	12.70(0.89)	&	79(6) & +13.66(0.86)	&	+24.90(0.48)	&	\b$-$4.4(1.1)	&	\b$-$8.5(0.7)	&	\b+7.7(0.7)	&	\b+0.7(0.9)	\\
216847	
&	6.76(0.67)	&	148(15) & +16.85(0.43)	&	+12.33(0.40)	&	\b$-$3.0(1.4)	&	$-$12.6(1.2)	&	\b+5.3(0.7)	&	\b$-$6.1(1.5)	\\
221756	
&	13.97(0.63)	&	72(3) & $-$17.71(0.45)	&	$-$45.77(0.42)	&	+13.4(0.5)	&	\b+7.7(0.5)	&	+10.9(0.6)	&	$-$16.7(0.6)		\\
\hline
\end{tabular}
\end{center}
\end{table*}

\begin{figure*}
\epsfxsize = 165mm
\epsffile{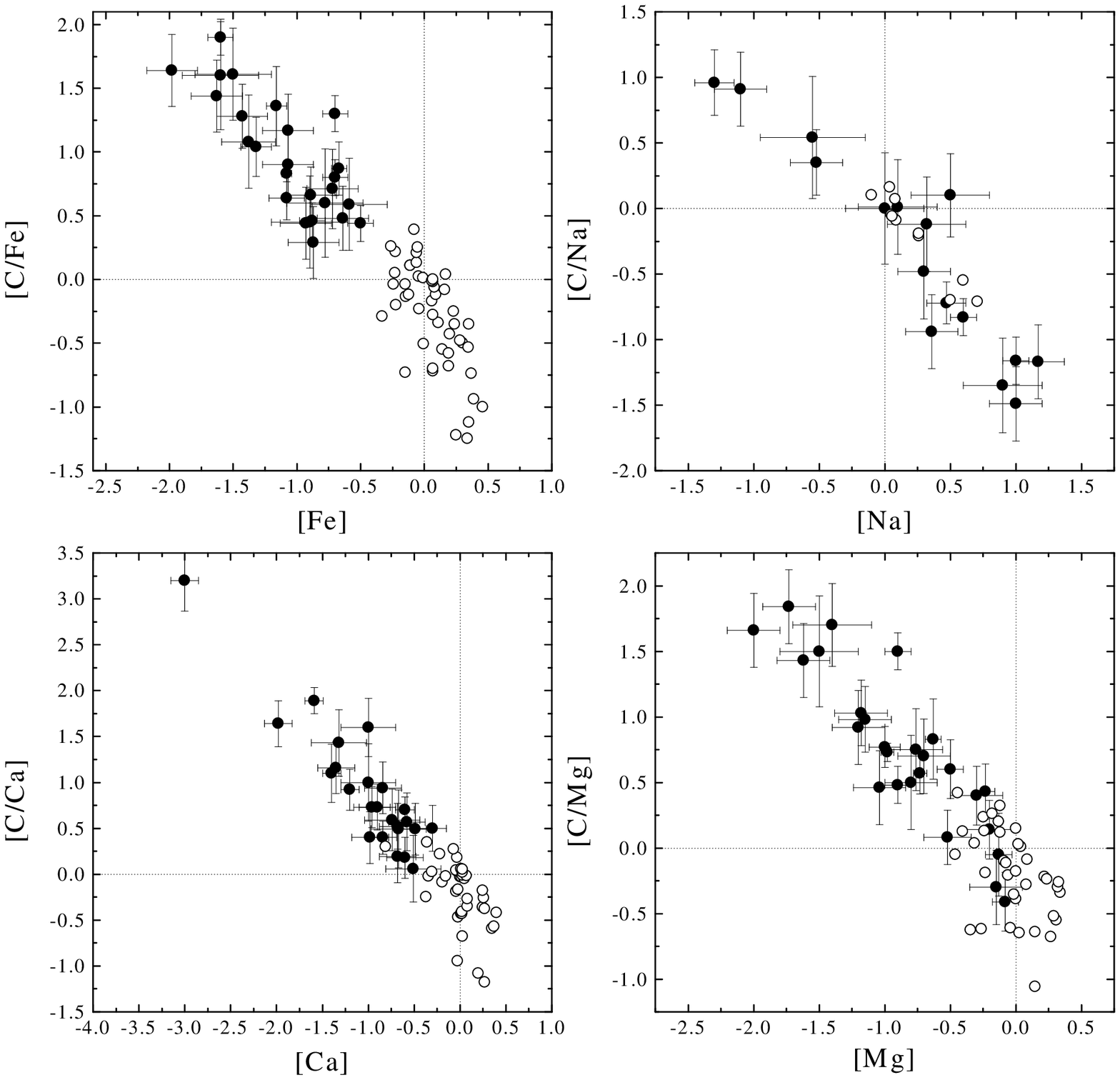}
\caption{Abundance ratios for the \LB stars (filled circles); open circles denote 
normal-type stars from the literature.} \label{abund}
\end{figure*}

\section{The abundance pattern of $\lambda$ Bo{\"o}tis stars} \label{abunds}

Since the first detailed abundance analysis in the late eighties,
there has been a question as to whether a unique abundance pattern exists 
for the \LB group. 
Most of the proposed candidates for membership have been found by
classification-resolution spectroscopy (typically 40\,\AA\,mm$^{-1}$ to
120\,\AA\,mm$^{-1}$). In the classical photographic domain (3800 to
4600\,\AA) spectral lines of Ca, Mg, Fe and Ti are the main contributors
to the overall appearance of the metallic-line spectrum. Gray (1988)
defined the membership of the class in terms of a weak Ca\,K and metallic-line
spectrum compared to the temperature classification of the hydrogen
lines. 

Paunzen et al.~(1999a) and Kamp et al.~(2001) already investigated the
behaviour of the lights elements C, N, O and S for a statistically significant
number of members:
\begin{itemize}
\item The star-to-star scatter for the abundances
of C, N, O and S is much smaller than that for heavier elements
\item The abundances of C, N, O and S are not strictly solar but range
from $-$0.8\,dex to +0.2\,dex compared to the Sun
\item Fe-peak elements are always more underabundant
\end{itemize}

Heiter (2002) and Heiter et al.~(2002) tried to shed more light on the abundance pattern
in the context of the theories proposed. They used mean values for a sample of
34 stars (not all elements' abundances were determined for all objects) and concluded: 
\begin{itemize}
\item The iron-peak elements from Sc to Fe as well as Mg, Si, Ca,
Sr and Ba are underabundant by 1\,dex compared to the Sun
\item Al is slightly more depleted whereas Ni, Y and Zr are slightly less
depleted
\item The mean abundance of Na is solar, but the star-to-star scatter
is about $\pm$1\,dex
\item The star-to-star scatter is twice as large as for a comparable sample
of normal stars
\end{itemize}
Otherwise they find no, or only a poor, correlation of individual abundances
with astrophysical parameters such as the effective temperature, surface
gravity, projected rotational velocity, age and pulsational period. 

We have used the individual abundances published by
Venn \& Lambert (1990), St\"{u}renburg (1993), Holweger \& Rentzsch-Holm (1995),
Chernyshova et al.\,(1998), Heiter et al.\,(1998), Paunzen et al.\,(1999a,b),
Kamp et al.\,(2001), Solano et al.\,(2001), Heiter (2002) and Andrievsky et
al.\,(2002) for members of the \LB group.
The individual values were weighted according to the errors listed in the references.
For our further analysis we have used only objects for which an abundance
of carbon or oxygen is available, since those are key elements for the definition
of the \LB group.

Values for stars of superficially normal type were taken from Adelman (1991, 1994,
1996), Adelman et al.\,(1991, 1997), Hill \& Landstreet (1993),
Hill (1995), Caliskan \& Adelman (1997) and Varenne \& Monier (1999).

Let us recall that the membership of an object in the \LB group is mainly based
on spectroscopy at classification resolution. The only other approach is the
definition of membership criteria in the UV region (Solano \& Paunzen 1999).
No reference in the literature was found which describes membership criteria
based on detailed abundance analysis in the optical region. If we compare
the status of other chemically peculiar stars of the upper main sequence then
a similar situation is evident (Wolff 1983; Cowley 1995). Objects have been
classified as being chemically peculiar but their individual elemental 
abundances differ widely. No attempt has so far been made to define the
membership of the classical CP stars to a sub-class by detailed abundances
alone (Preston 1974).

The \LB group is unusual in this respect since it shows strong
underabundances, not found for any other group, of most heavier elements.
The only exceptions are intermediate and true
Population\,II-type objects and field blue stragglers,
post-AGB and F-weak stars (Gray 1988, 1989; Jaschek, Andrillat \& Jaschek
1989; Andrievsky, Chernyshova \& Ivashchenko 1995; van Winckel, Waelkens \& Waters 1995). 
Objects with very low surface
gravities (post-AGB and Population\,II-type objects) are easily distinguished
even at classification resolution and will not be considered in the
following discussion. For the other groups the underabundances
of the Fe-peak elements are rather moderate. But, more importantly, the abundances of the light
elements C, N, O and S scale just like those of the heavier elements.
We have therefore chosen to use [C/Z] versus [Z] diagrams ([O/Z], [N/Z] and [S/Z] 
behave analogously) in order to investigate the behaviour of the group of \LB stars. 
The field blue stragglers, intermediate Population\,II and F-weak types of
stars all fall in the area around [C/Z]\,$\approx$\,0 in these diagrams.

Figure \ref{abund} shows such diagrams for the elements Fe, Na, Ca and Mg
for the stars on our program as well as for those of superficially normal types. From this
Figure we are able to conclude:
\begin{itemize}
\item All program stars are Population\,I-type objects with [C/Fe]\,$\gg$\,0
\item The \LB stars exhibit iron abundances that are significantly
lower than those found for the superficially normal stars
\item There is a large overlap for all other heavier elements 
\end{itemize}
Cowley et al.~(1982) have proposed that \LB\hskip-2pt-type and other weak-line stars
may arise from small ($\approx$\,0.3\,dex) abundance fluctuations in the interstellar
medium. That might be true for a small fraction of the objects, but underabundances up
to a factor of 100 can not be explained without some other mechanism such as diffusion,
accretion or mass-loss.

\begin{figure}
\begin{center}
\epsfxsize = 74mm
\epsffile{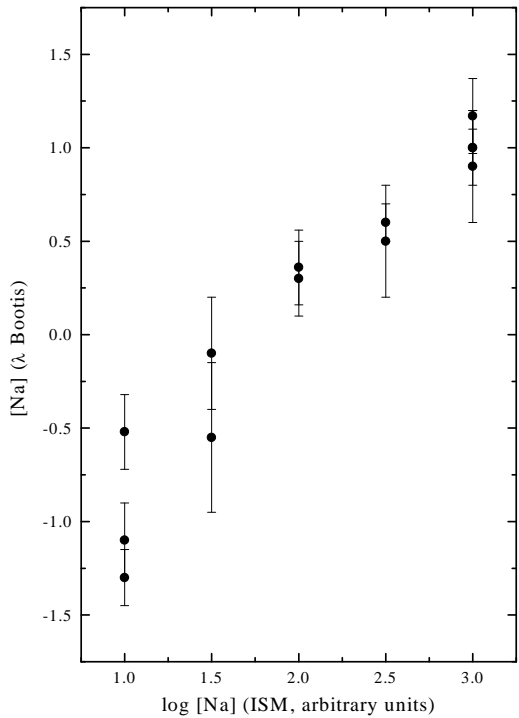}
\caption{Correlation between the sodium abundances for 13 \LB
stars and the values, taken from Welsh et al.\,(1998), 
for the surrounding environment. 
The latter values are scaled in three steps (a value
of three denotes the highest density).} \label{nad}
\end{center}
\end{figure}

A very intriguing fact is that eight program stars exhibit magnesium abundances
which range from $-$0.52 to $-$0.13\,dex, which seems to contradict the classification
based on moderate-resolution spectroscopy. One of the most important classification
criterion is the moderate to extreme weakness of the Mg\,II\,4481\,\AA\,line (Gray 1988). 
What can actually cause that discrepancy? Slettebak, Kuzma \& Collins (1980) have shown that the
equivalent width of the line decreases with increasing rotation for models later than
A0. The same fact was also described by Abt \& Morrell (1995). 
A much stronger effect was found for H$\gamma$. That means
that a rapidly rotating star will be classified much later according to
H$\gamma$ than on the basis of
Mg\,II\,4481\,\AA. Taking a rapidly rotating A5 type star one would classify
it as hF1mA7 or metal-weak. However, the three objects with the highest magnesium
abundances of our sample are indeed the fastest rotators: HD~193256 (250\,kms$^{-1}$;
$-$0.15\,dex), HD~170680 (205\,kms$^{-1}$; $-$0.20\,dex) and HD~198160 (200\,kms$^{-1}$;
$-$0.13\,dex). But there is no overall correlation for the whole sample, i.e.
fast rotators do also exhibit rather strong underabundances (e.g. HD~111604) and vice
versa. Otherwise the fast rotators are not outstanding in any respect.

One observational fact is the wide range of abundances ($-$1.3 to +1.2\,dex)
for sodium (Table \ref{lb_abund} and Fig. \ref{abund}) first indicated in the work of 
St\"urenburg (1993). 
In the recent literature no explanation of the variability has been given. 
Sodium is the only element for which such a behaviour has been detected so far. 
Furthermore, predictions for that element 
have not been discussed so far within any proposed theory (Turcotte \& Charbonneau 1993).
In an effort to shed some light on the subject,
we have investigated whether a correlation can be found between the individual sodium abundances and the
density of sodium in the surrounding interstellar medium. 
Welsh, Crifo \& Lallement (1998) published
the local distribution of interstellar Na\,I within 250\,pc of the Sun. They
used all published absorption densities and the distances derived from the Hipparcos satellite.
The densities were scaled in three steps (a value of three denotes
the highest density) and plotted for different Galactic 
coordinates and distances from the Sun. We have selected members of the \LB group
whose sodium abundances are known and for which nearby data points are available in the maps 
of Welsh et al.\, (1998). We have also checked more recent 
references such as Sfeir et al.\,(1999) and Vergely et al.\,(2001) which give
essentially the same results. In total, thirteen objects fulfill the requirements: HD~319,
HD~31295, HD~74873, HD~84123, HD~125162, HD~183324, HD~192640, HD~193256, HD~193281,
HD~198160, HD~204041, HD~210111 and HD~221756. 

We have to emphasize that not a single sodium abundance for the line of sight
to any \LB star has yet been measured. Bohlender et al.\,(1999) reported a few
detections of such features, e.g. for HD~319, HD~192640 and HD~221756, but they did
not derive quantitative densities or abundances.

Figure \ref{nad} shows the correlation of the individual sodium abundances with
the absorption densities for the local interstellar medium (ISM hereafter). 
Since there is a linear
correlation visible it suggests that there is an interaction (e.g. accretion) 
between the stars and their environments at some stage of
stellar evolution. The correlation does not reflect any
age dependency, since objects are included with 0.33\,$<$\,$t_{\rm rel}$\,$<$\,1.01.
This rather small sample is reasonably representative of the whole sample
of program stars in terms of its distribution in $t_{\rm rel}$ and effective temperature.
Unfortunately it is not possible to include any data for superficially normal
objects in Fig. \ref{nad} since no sodium abundances for bright field
stars are available. The only data are those published by
Varenne \& Monier (1999; Fig. \ref{abund}) for members of the Hyades.

\begin{figure}
\begin{center}
\epsfxsize = 74mm
\epsffile{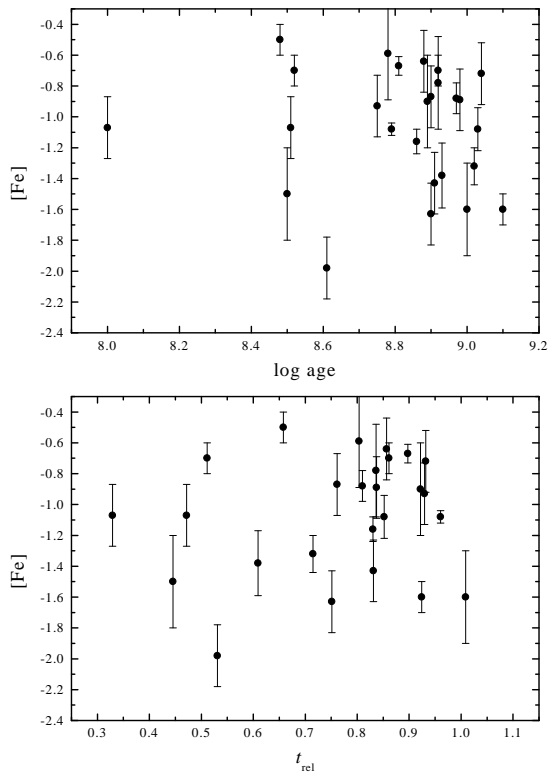}
\caption{No correlation is found between iron abundance and age.} \label{felogage}
\end{center}
\end{figure}

From our current analysis two main questions arise for the \LB phenomenon:
How many mechanisms are involved? What are the observational constraints?
It is clear that there is one mechanism which produces the observed pattern throughout 
the whole 
MS lifetime for stars between late B and early F types. We are not able to decide
if it is `internal' or `external' but we have some hints about it. 
It is highly improbable that one mechanism works for early evolutionary
stages and an independent second one at very late stages
producing the same absolute abundance pattern. That seems to be supported by the 
non-existence of a correlation between the iron abundance and age (Fig. \ref{felogage}).
The abundance of sodium for the stellar atmosphere is correlated with that of
the local ISM. There are two possible explanations for that: 1) the atmospheric
abundance resembles the one from the cloud in which
the star was born, or 2) it currently interacts with the local ISM. If we
believe in the first interpretation then all other abundances ought also to
resemble those in the local 
ISM. That seems not to be the case, since many other `normal' stars located within 
the same ISM clouds show no significant elemental underabundances at all. So
we are left with the picture of an interaction between the star and its environment.

\begin{figure}
\begin{center}
\epsfxsize = 74mm
\epsffile{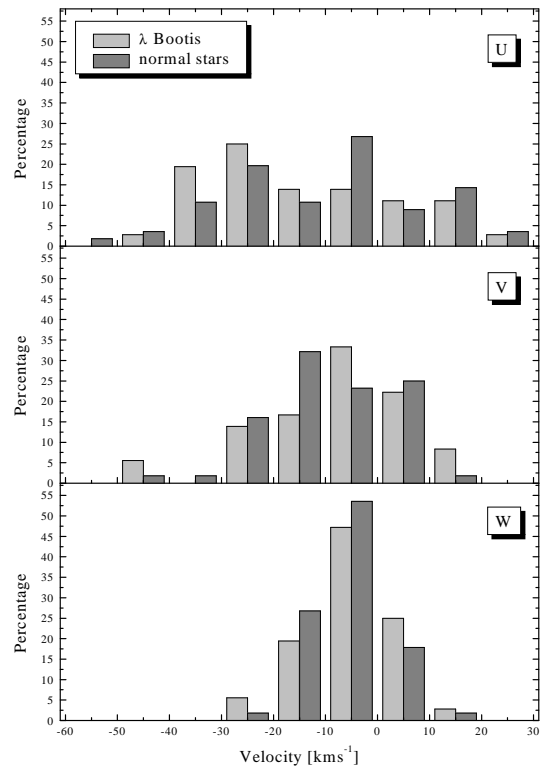}
\caption{The histograms of the Galactic space-velocity components ($U, V$ and $W$)
for the \LB and normal-type samples. The distributions agree at significance
levels of 72\% ($U$), 74\% ($V$) and 97\% ($W$).} \label{motion}
\end{center}
\end{figure}

\section{The space motions of $\lambda$ Bo\"otis stars}

In order to study the kinematics of nearby stars, one needs to calculate
the Galactic space-velocity components ($U$, $V$ and $W$), given the
star's proper motion, parallax (both listed in the Hipparcos catalogue) and
radial velocity. 
Here, the formulae and error propagation are taken
from the Hipparcos documentation as well as from Johnson \& Soderblom
(1987).

It was chosen to calculate heliocentric Galactic velocity components,
which can easily be corrected for the solar motion. A right-handed
coordinate system for $U$, $V$ and $W$ was used, so they are positive
in the directions of the Galactic centre, Galactic rotation and
the North Galactic Pole. The `standard solar motion' in this system
would be (+10.4, +14.8, +7.3) taken from Mihalas \& Routly (1968).

The authors followed the approach of the Hipparcos consortium who discussed
the calculation, transformation matrix and error propagation.
We would like to give only a short description of the error estimation.
The general equation for the variance of a function of several variables is
used for this purpose. 
That formula is true only if the covariances are zero (the errors are
uncorrelated). That assumption is fulfilled since the two proper-motion
components are measured independently. Furthermore it is assumed that
only the radial velocity, proper motions and parallax 
contribute to the error distribution. For nearby stars (about 25\,pc) an error
of the proper motions of about $\pm$3\,mas\,yr$^{-1}$
(most of the Hipparcos measurements are a factor three better
than that) corresponds to an uncertainty in the transverse motion of 
only about 0.5\,km\,s$^{-1}$; it is clear that the errors of 
the radial velocities are much more significant.

The radial velocities for our program stars are
listed in Table \ref{lb_mot}. The following references were used for
that Table: Wilson (1953), Evans (1967), Batten, Fletcher \& MacCarthy (1989), Barbier-Brossat,
Petit \& Figon (1994), Duflot et al.\,(1995), Hauck et al.\,(1995, 1998), 
Fehrenbach et al.\,(1996, 1997), 
Nordstr\"om et al.\,(1997) and Grenier et al.\,(1999a,b). 
The mean and the error of the mean were calculated without weighting the
individual measurements.
From Table \ref{lb_mot} only three stars (HD~319, HD~110411 and
HD~204041) show evidence for variable radial velocities
but no prominent photometric or spectroscopic variability has been detected so far. 
Moreover, Table \ref{lb_mot} lists the kinematic data for the program \LB stars
from the Hipparcos database. The investigated sample is limited
in distance ($\approx$\,220\,pc) because of the lack of Hipparcos data and/or
radial velocities for more distant
objects. No star exceeds $\mid$$V$$\mid$\,=\,50\,km\,s$^{-1}$ and the
sample is very homogeneously distributed. A similar conclusion is given by
G\'omez et al.\,(1998a) for classical CP stars as well as by G\'omez et al.\,(1998b)
for a small sample of \LB stars. 

We have compared the velocities of stars in our \LB sample with the results of Caloi et
al.\,(1999). They studied the relationship between the kinematics, age and 
heavy-element content for the solar neighbourhood, using the Hipparcos data. For
$V$\,$<$\,$-$40\,km\,s$^{-1}$, $\mid$\,$U$\,$\mid$\,$>$\,+60\,km\,s$^{-1}$ and 
$\mid$\,$W$\,$\mid$\,$>$\,+30\,km\,s$^{-1}$, the minimum stellar age is about 2\,Gyr.
Only two of the \LB stars would meet the $V$ criterion, HD~6870
($V$\,$=$\,$-$41.7(2.8)\,km\,s$^{-1}$) and HD~84123 ($V$\,$=$\,$-$45.4(4.4)\,km\,s$^{-1}$),
but their $U$ and $W$ velocities (like those of all other stars in
the sample) are well below the limits given. Overall, the velocities 
are typical of true Population\,I objects.

The comparison with the sample of normal stars (Fig. \ref{motion})
shows very good agreement between the two distributions. They agree at significance
levels of 72\% ($U$), 74\% ($V$) and 97\% ($W$).

Notice that Faraggiana \& Bonifacio (1999) made a similar analysis for a different
sample of candidate \LB stars (see Section 4 therein). They only
investigated the parameter $(U^2+W^2)^{1/2}$ which is a measure of the kinematic energy
not associated with Galactic rotation. They come to the conclusion that all
of their program stars are qualified as disk members. Since they give no individual values 
for the space motions
and radial velocities, we are not able to compare our results with theirs.

\section{Conclusions}

We have used all currently available photometric data as well as Hipparcos data to determine
astrophysical parameters such as the effective temperatures, surface gravities and
luminosities. As a next step, masses and ages were calibrated within appropriate 
post-MS evolutionary models. Furthermore, Galactic space motions were calculated
with the help of radial velocities from the literature. The comparison with already
published results shows good agreement of the derived parameters. 
All results were compared with those of a test sample of normal-type
objects in the same spectral range chosen in order to match the $(B-V)_0$
distribution of the \LB group. 
From a comprehensive statistical analysis we conclude:
\begin{itemize}
\item The standard photometric calibrations within the Johnson {\it UBV}, Str\"omgren $uvby\beta$
and Geneva 7-colour systems are valid for this group of chemically peculiar stars.
\item The group of \LB stars consists of true Population\,I objects which can be found
over the whole area of the MS with a peak at a rather evolved stage ($\approx$\,1\,Gyr).
That is in line with the distribution of the test sample.
\item The \LB type group is not significantly distinct from normal
stars except, possibly, by having slightly lower temperatures and masses for
$t_{\rm rel}$\,$>$\,0.8. The $v$\,sin\,$i$ range is rather narrow throughout the
MS with a mean value of about 120\,kms$^{-1}$.
\item There seems to exist a non-uniform distribution of effective temperatures for
group members with a large proportion of objects (more than 70\%)
cooler than 8000\,K.
\item It seems that objects with peculiar hydrogen-line profiles are preferentially 
found among later stages of stellar evolution.
\item No correlation of age with elemental abundance or projected rotational
velocity has been detected.
\item A comparison of the stellar Na abundances with nearby IS sight lines hints 
at an interaction between the \LB stars and the ISM.
\item There is one single mechanism responsible for the observed phenomenon which produces 
moderate to strong underabundances working continuously from very early (10\,Myr) to very late
evolutionary stages (2.5\,Gyr). It produces the same absolute abundances throughout the MS lifetime
for 2\% of all luminosity class V objects with effective temperatures from 10500\,K
to 6500\,K. 
\item The current list of stars seem to define a very homogeneous group, validating
the proposed membership criteria in the optical and UV region.
\end{itemize}
These rather strict observational results for a significant number of \LB stars
will need to be taken into account in future work on theories
and models trying to explain the phenomenon. The constraints presented here
will help considerably to reduce the number of free parameters in the models and
finally to provide a critical test for them.
\\
\\
{\noindent \footnotesize {\bf Acknowledgements.}
This work was partly supported by the Fonds zur F\"orderung der
wissenschaftlichen Forschung, project P14984 and it is dedicated to
G.~Derka who died during its preparation. Use was made of the 
SIMBAD database, operated at CDS, Strasbourg, France and
the GCPD database, operated at the Institute of Astronomy of the University
of Lausanne.
}

\end{document}